\documentclass[12pt, preprint]{aastex}

\newcommand{\K}{\,{\rm K}}

\usepackage{lscape}
\usepackage{longtable}

\shorttitle{MWC275 and AB~Aurigae disk models}
\shortauthors{Tannirkulam et al.}

\begin{document}




\title{A Tale of Two Herbig Ae stars -MWC275 and AB Aurigae: Comprehensive Models for SED and Interferometry.}



\author{A.~Tannirkulam\altaffilmark{1},
 J.~D.~Monnier\altaffilmark{1}, T.~J.~Harries\altaffilmark{2},  R. Millan-Gabet\altaffilmark{3},  
 Z. Zhu\altaffilmark{1}, E. Pedretti\altaffilmark{4},  M. ~Ireland\altaffilmark{5}, P. Tuthill\altaffilmark{5}, T. ~ten ~Brummelaar\altaffilmark{6}, H. McAlister\altaffilmark{6}, C. Farrington\altaffilmark{6}, P.~J.~Goldfinger\altaffilmark{6}, J. Sturmann\altaffilmark{6}, L. Sturmann\altaffilmark{6},  N. Turner\altaffilmark{6}
 }
\altaffiltext{1}{atannirk@umich.edu: University of Michigan, Astronomy Dept, 
500 Church Street, 1017 Dennison Bldg, Ann Arbor, MI 48109-1042, USA}
\altaffiltext{2}{University of Exeter, School of Physics, Stocker Road, Exeter, EX4 4QL, UK}
\altaffiltext{3}{Michelson Science Center, Pasadena, CA, USA}
\altaffiltext{4}{University of St. Andrews, Scotland, UK}
\altaffiltext{5}{School of Physics, Sydney University, N. S. W. 2006, Australia}
\altaffiltext{6}{CHARA, Georgia State University, Atlanta, GA, USA}


\begin{abstract}
\noindent   We present comprehensive models for the Herbig Ae stars MWC275 and AB~Aur that aim to      
explain their spectral energy distribution (from UV to millimeter) and long baseline interferometry (from near-infrared to millimeter) simultaneously. Data from the literature, combined with new mid-infrared (MIR) interferometry from the Keck Segment Tilting Experiment, are modeled using an axisymmetric Monte Carlo radiative transfer code. Models in which most of the near-infrared (NIR) emission arises from a dust rim  fail to fit the NIR spectral energy distribution (SED) and  sub-milli-arcsecond NIR CHARA interferometry. Following recent work, we include an additional gas emission component with  similar size scale to the dust rim,  inside the sublimation radius, to fit the NIR SED and long-baseline  NIR interferometry on MWC275 and AB~Aur. In the absence of shielding of star light by gas, we  show that the gas-dust transition region in these YSOs will have to contain highly refractory dust, sublimating at   $\sim$1850K. Despite having nearly identical structure in the thermal NIR, the outer disks of MWC275 and AB~Aur differ substantially. In contrast to the AB~Aur disk, MWC275 lacks small grains in the disk atmosphere capable of producing significant 10-20 $\mu$m emission beyond $\sim$7AU, forcing the outer regions into the ``shadow'' of the inner disk.
\end{abstract}




\section{Introduction}
\label{intro}
Herbig Ae (HAe) stars are pre-main-sequence stars of intermediate mass (1.5-3 solar masses). They exhibit a robust excess in emission over stellar photospheric values
 from near-infrared (NIR) to the millimeter (mm) wavelengths. This excess is now attributed to the passive reprocessing of stellar light by dust in the circumstellar environment \citep{lkha2001, natta2001, dullemond2001}. The geometry of the circumstellar environment of HAe stars has been actively debated in the astronomy community over the last two decades. Some of the early workers  in this field \citep{Hillen92} showed that spectral energy distribution (SED) of  HAe stars could be explained by emission from circumstellar matter in disk-like geometry. Others \citep{Mirosh97} argued that the emission could also arise from dust in a spherical geometry around the star, proving the inadequacy of SED modeling alone in uniquely fixing the geometry of the circumstellar matter. The first observational evidence in favor of a disk geometry came from millimeter (mm) interferometry in the form of asymmetries detected \citep{Man97} in the mm images. Asymmetries in the NIR emission were also detected by the Palomar Test-Bed Interferometer \citep{eisner2003, eisner2004}, settling the debate in support of a disk geometry for circumstellar material in Herbig Ae stars.
 
 Most interferometric studies of HAe stars have relied on simple geometric models \citep{Man97, rmg1999, rmg2001, eisner2003, eisner2004, monnier2005}  that explain the emission geometry of the system in only narrow wavelength ranges. This method, albeit extremely useful in elucidating some of the morphology details, is not adequate for exploring the interdependency in structure of the inner and outer parts of the disk. A number of studies \citep{dullemond2001, dullemond2004, boekel2005b} have shown that the structure of the inner disk at fractions of an AU scale clearly affects the structure of the outer disk. A complete understanding of the circumstellar disk structure in HAe stars therefore requires models that simultaneously explain the SED and interferometry over a large wavelength range. Such models have begun to appear in the literature only recently \citep{ponto2007, kraus2007}.

 In this paper, we develop comprehensive disk models to  explain the SED and interferometry of the HAe stars MWC275 and AB Aur.   MWC275 and AB~Aur are prototype pre-main-sequence stars of similar ages and  spectral type  with extensive circumstellar disks. Due to the availability of photometric and interferometric data over a large wavelength range, MWC275 and AB~Aur are ideal candidates for testing disk models  for YSOs. The extent  of their circumstellar-dust disks was first measured by \citet{Man97} to be several 100AU using the Owens Valley Radio Observatory (OWRO). \citet{Natta2004} resolved the MWC275 disk in the mm and reported a de-convolved,  projected dust-disk size of 300AU$\times$180AU.  More recently, \citet{isella2007} analyzed IRAM, SMA and VLA continuum and $^{12}$CO, $^{13}$COand $^{18}$CO line data  constraining the gas-disk radius to be 540AU with the gas in Keplerian rotation around the central star. Scattered light studies of MWC275 \citep{grady2000} and AB~Aur \citep{grady1999, oppenheimer2008} show the presence of arcs and rings in the circumstellar disk.  \citet{corder2005} resolved the AB~Aur CO disk radius to be $\sim$600AU, finding strong evidence for Keplerian rotation for the bulk of the disk.  \citet{corder2005} and \citet{lin2006} detected spiral arms in CO emission with radii of  $\sim$150AU, while \citet{fukagawa2004} detected similar structure in Subaru H-band scattered light images. AB~Aur also has substantial envelope material on scales larger then 600 AU \citep{grady1999, semenov2005, corder2005, lin2006}.
 
 MIR emission probes the giant planet  formation region in circumstellar disks \citep{calvet1992, cg97, dullemond2001} with the emission arising from warm dust (T $>$ 150K). \citet{Meeus2001} and \citet{boekel2005} used the 10$\mu$m MIR silicate emission feature from MWC275 and AB~Aur to show that dust grains in these systems had  grown larger than the typical interstellar medium grain sizes. \citet{marinas2006} imaged AB~Aur  at 11.7$\mu$m and found  the  emission FWHM size to be 17$\pm$4 AU consistent with the flared disk models of \citet{dullemond2004}.  In this paper,  we present  new 10$\mu$m  measurements of AB~Aur  and MWC275 with the Keck Segment Tilting Experiment \citep[described in \S\ref{data}]{Monnier2004}.   In contrast to AB~Aur, the MWC275 disk is unresolved by the Segment Tilting Experiment (maximum baseline of 10m),  requiring the VLT Interferometer (100m baseline) to probe to its MIR  structure \citep{Leinert2004}. These observations suggest that  MWC275 disk differs considerably from AB~Aur and we present  a detailed comparison of the two disk structures in the discussion  (\S\ref{discuss}). 
 
 Thermal NIR emission probes hot regions (typically the inner AU) of the disk with temperatures greater than 700K. The NIR disks of MWC275 and AB~Aur were first resolved with IOTA by \cite{rmg1999, rmg2001} and subsequently observed at higher resolution with PTI \citep{eisner2004}, Keck Interferometer \citep{monnier2005} and the CHARA  interferometer array \citep{Tannirk2008}. In \citet[hereafter T08]{Tannirk2008} we showed that inner-disk models in which majority of the K-band emission arises in a dust rim \citep{dullemond2001, isella2005, Tannirk2007} fail to fit the CHARA data at milli-arcsecond resolution. We also demonstrated that the presence of additional NIR emission (presumably from hot gas) inside the dust destruction radius can help explain the CHARA data and the NIR SED. First calculations for the effects of gas on  rim  structure \citep{muzerolle2004}  showed that for plausible disk parameters,  presence of gas does not modify dust-rim geometry significantly. Besides a poorly understood interferometric visibility profile, MWC275 also displays as yet ill-understood NIR and MIR SED time variability \citep{sitko2007} which has been interpreted as variations of the inner disk structure. In  \S\ref{MWC275_model} and \S\ref{ABAur_model} we present a detailed analysis of  the NIR visibility and SED for MWC275 and AB~Aur, placing constraints on the wavelength dependence of the opacity source inside the dust destruction radius.

 In this study, we focus on (i) explaining the inner-disk structure and discuss important open problems and  (ii) modeling the MIR emission morphology of the disks and the shape of the MIR spectrum. The paper is organized into 7 sections with \S\ref{data} detailing the observations. \S\ref{model} explains the disk model and the modeling strategy. \S\ref{MWC275_model} and \S\ref{ABAur_model} analyze MWC275 and AB~Aur SED and visibilities in relation to the  disk models. We present a discussion on our results  and our conclusions  in \S\ref{discuss} and \S\ref{conclusion} respectively.

\section{New Observations and Literature Data}
\label{data}
The properties of the circumstellar disks in MWC275 and AB Aur have been  constrained using IR and mm interferometry  and SED.  We include new  NIR interferometry from the CHARA array, MIR interferometry from the Keck Segment Tilting Experiment and mm interferometry from the literature in our analysis. Optical and NIR photometry obtained at the MDM Observatories, and mid and far-infrared data from ISO are also included to constrain model SED. We describe the data in detail in the following paragraphs.

\begin{figure}[h]
\begin{center}
{
\includegraphics[angle=90,width=5.in]{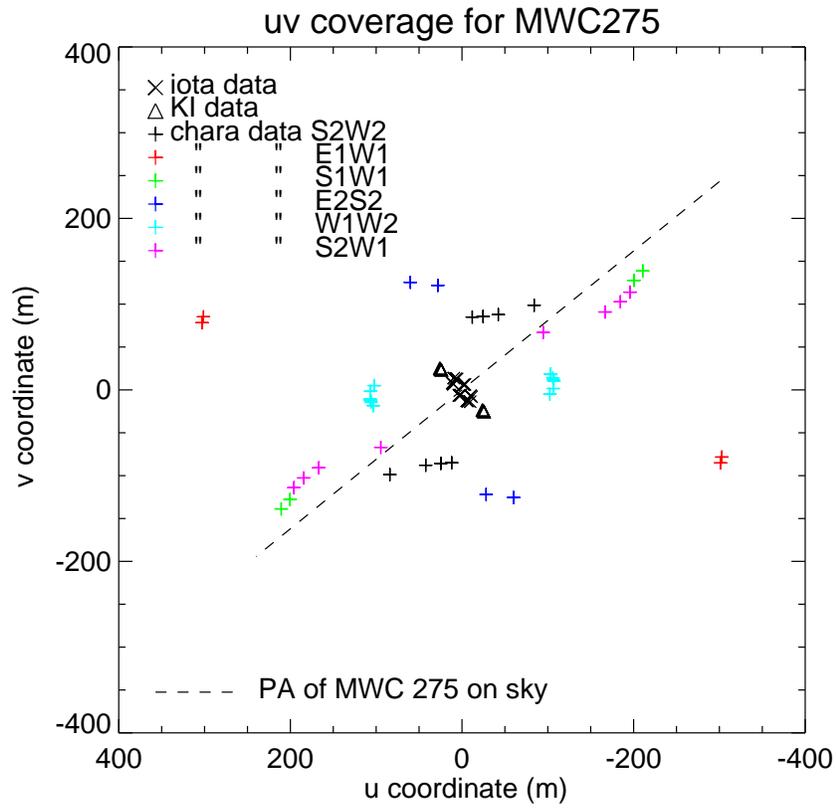}
}

\caption{uv coverage for MWC275. We include data
from  KI \citep{monnier2005}, IOTA \citep{monnier2006} and CHARA \citep{Tannirk2008} in our analysis. A position angle (measured East of North) of  136$^o$ for  MWC275 is marked in the left panel.
}
\label{uv_cover_MWC275}
\end{center}
\end{figure}

\begin{figure}[h]
\begin{center}
{
\includegraphics[angle=90,width=5.in]{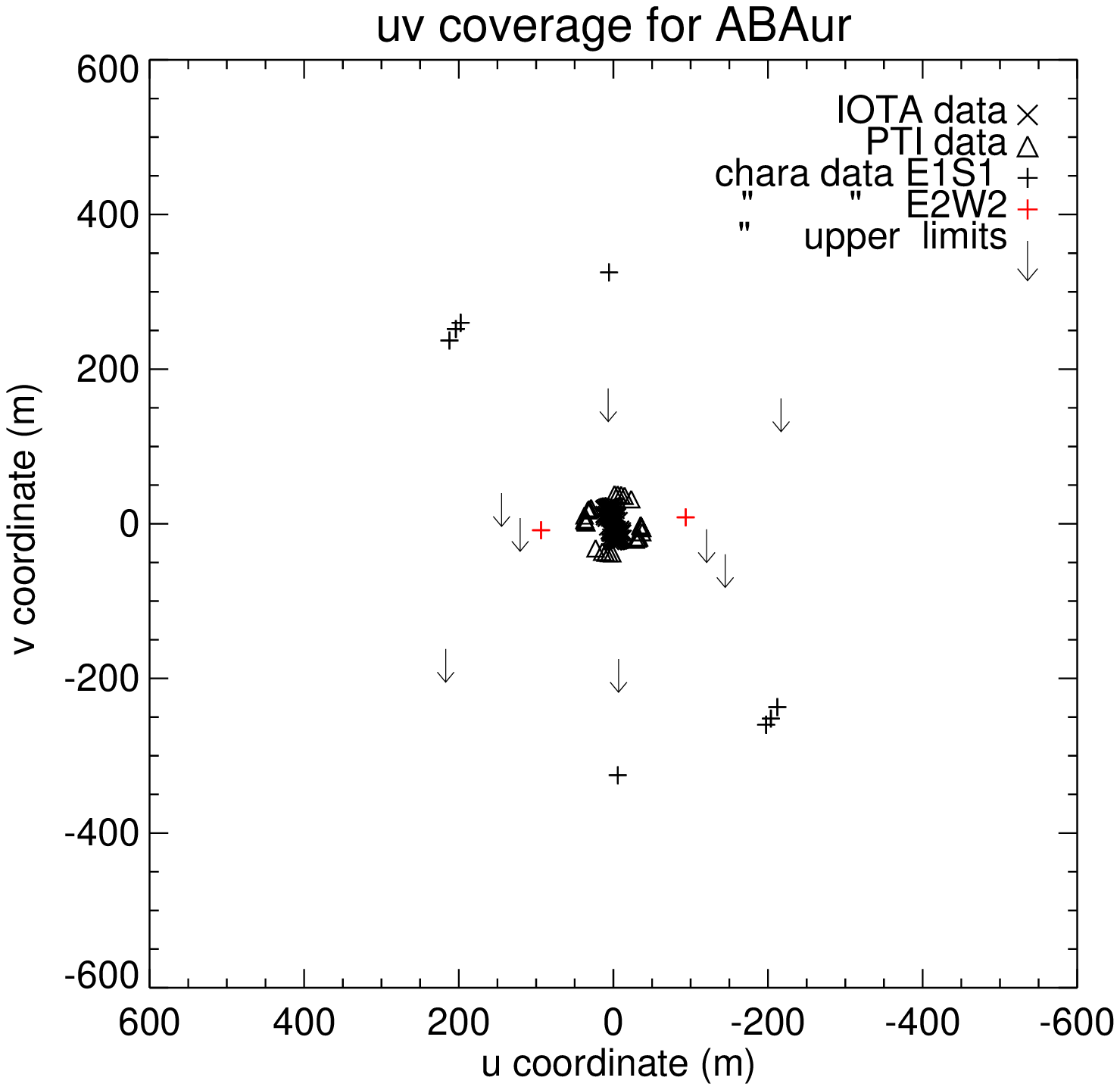}
}

\caption{uv coverage for AB~Aur. We include data
from  PTI \citep{eisner2004},  IOTA  and CHARA \citep{Tannirk2008} in our analysis.}
\label{uv_cover_ABAur}
\end{center}
\end{figure}

K-band (central wavelength of 2.13$\mu$m, $\Delta\lambda$ 0.3$\mu$m) interferometry on MWC275 and AB~ Aur, first presented in T08, was obtained using the CHARA array with the 2-beam CHARA ``Classic''  combiner at the back end \citep{ten2005}.  The targets  were observed with multiple baselines of CHARA at a variety of orientations in 8  runs between June 2004 and June 2007. The longest baseline observation for MWC275 was 325m (resolution \footnote {Resolution is defined as $\frac{\lambda}{2D}$, where $\lambda$ is wavelength of observation and D is the interferometer baseline length.}  of 0.67 milli-arcseconds) and 320m (resolution of 0.68 milli-arcseconds) for AB~Aur. The data were reduced using standard CHARA reduction software \citep{ten2005} and these results were cross-checked with an independent code developed at University of Michigan. HD164031 , HD166295  and HD156365 with uniform-disk (UD) diameters of 0.83$\pm$ 0.08 mas, 1.274$\pm$0.018 mas and 0.44$\pm$0.06 mas \citep[and getCal - http://mscweb.ipac.caltech.edu/gcWeb/gcWeb.jsp]{merand}  were used as calibrators for MWC275. AB~Aur visibilities were calibrated with HD29645 (UD diameter=0.54$\pm$0.07 mas, getCal) and HD31233 (UD diameter=0.76$\pm$0.13 mas, getCal). During the reduction procedure the flux ratios between the two interferometer telescopes were carefully monitored for the source and the calibrators. Data points having a flux ratio variation greater than 10\%  of the mean, indicating large changes in instrument alignment, were discarded. Three MWC275 data points were removed by this criterion.  The procedure was adopted to minimize calibration errors caused by differences in beam overlap in the source and calibrator.  The relative visibility errors  which include calibration errors, statistics and uncertainties in calibrator size,  are at the $\sim$6\% level, typical for CHARA Classic. In T08, we had shown the reduced data briefly in the form of visibility interferometric-baseline plots. Here we present  the complete observing logs with the uv coverage (see Figs \ref{uv_cover_MWC275} and \ref{uv_cover_ABAur}) and the calibrated visibilities listed in Tables \ref{uv_table1} and \ref{uv_table2}.  NIR interferometry from IOTA \citep{monnier2006}, PTI \citep{eisner2003, eisner2004} and the Keck Interferometer \citep{monnier2005} have also been included to constrain the models.

 MW275 and AB~Aur were observed with the Keck Segment Tilting Experiment \citep{Monnier2004}  to study their MIR emission morphology. Four subsets of Keck primary mirror segments were repointed  and rephased so as to achieve four separate sparse aperture Fizeau interferometers, each with a separate pointing origin on the LWS detector \citep{Jones1993}.  The details of the experiment and the data reduction procedure are  provided in \citet{Monnier2004} and Monnier et al. (2008, in prep).  The calibrated  data are listed in Table \ref{LWS}. In addition to the Keck Segment Tilting data,  we also include MWC275 MIR intereferometry from VLTI-MIDI \citep{Leinert2004} in our analysis.
 
 Millimeter  interferometry from  \citet{Man97}, \citet{Natta2004},  \citet{semenov2005}, \citet{lin2006} and \citet{isella2007} are used to constrain the circumstellar disk masses and disk sizes and surface density profile. In conjunction with the interferometry, the shape of the mm SED provides information on sizes of the dust grains in the bulk of the circumstellar disks.
   
To constrain the SED computed from radiative transfer models we obtained optical and NIR photometry on MWC275 and AB~Aur with the MDM 2.4m telescope at Kitt Peak. Due to the  high brightness of our targets,  we defocussed the telescope during observations to avoid camera saturation. After bias correction, flat fielding, and background subtraction, the reduced images were used to obtain source counts using ATV - an aperture photometry code \citep{barth2001}. Targets were calibrated using Landoldt standards \citep{landolt1983}. The  calibrated UBVRIJHK  photometry are listed in Tables \ref{UBVR} and \ref{IJHK} in the Appendix. We also include photometry for a number of other YSOs that we observed.  NIR photometry  from \cite{sitko2007} , mid and far-IR SED from \cite{Meeus2001} and mm fluxes  \citep{Man97, Natta2004, semenov2005, lin2006} were also  used in the model analysis. \citet{boekel2005} modeled the 10$\mu$m spectra of a large sample of  Herbig Ae stars and derived the mineralogy and typical grain sizes responsible for the emission. We use results from   \citet{boekel2005} for fixing the composition of dust grains in the atmosphere of the MWC275 and AB~Aur disks.  

A compilation of stellar properties and circumstellar disk properties extracted from the literature is listed in Tables \ref{photometry_a}, \ref{photometry_b},  \ref{MWC275_properties} and \ref{ABAur_properties}.

\section{Circumstellar Disk Modeling}
\label{model}
The circumstellar material around MWC275 and AB~Aur is modeled as a passive  dust disk reprocessing stellar radiation \citep{dullemond2001}. The disk is in thermal and hydrostatic equilibrium and can be divided into 3 distinct regions (Fig \ref{schematic}) -

\begin{list}{$\bullet$}{\itemsep=0in}
\item{\bf{Curved Inner Rim}} Sufficiently close to the star (distance depends on stellar luminosity and dust sublimation temperatures), dust in the circumstellar disk reaches sublimation temperatures and evaporates. Inside of the evaporation radius,  the disk is optically thin. The truncated dust disk  is frontally illuminated by the star and forms a `rim' whose shape depends sensitively on dust properties \citep{isella2005, Tannirk2007}. The rim puffs up, traps a significant fraction of stellar photons and re-radiates predominantly in the NIR \citep{dullemond2001}. 

\item{\bf{Shadow Region}} The inner rim casts a geometric shadow on the region behind it \citep{dullemond2001, dullemond2004}, preventing it from receiving direct star light. The shadow is heated by scattered photons from the rim edge and through diffusion. The size of the shadow depends on the rim geometry, mass of dust in the outer disk and dust grain properties in the outer disk.

\item{\bf{Flared Disk}} If sufficient dust is present in the outer disk, the disk eventually emerges from the shadow and ``flares''. The flared disk emits radiation in the MIR and longer wavelengths.
\end{list}
Details of the modeling procedure and comparison to data are described below.

\subsection{The Monte Carlo Transfer Code - TORUS}
\label{sublimation_model}
The calculations in this paper were performed using the {\sc torus}
Monte-Carlo radiative-transfer code \citep{harries2000, harries2004,
kurosawa2004, Tannirk2007}. Radiative equilibrium is computed using Lucy's
\citep{lucy1999} algorithm on a two-dimensional, cylindrical
adaptive-mesh grid. The initial density structure for the disk calculations is based on
the canonical description of the $\alpha$-disk developed by
\citet{shakura1973}, viz
\begin{equation}
\rho(r,z) = \rho_0 \left (\frac{r}{r_0}\right)^{-\alpha}\exp\left[ -\frac{1}{2}\frac{z^2}{h(r)^2}\right]
\end{equation}
where $r$ is the radial distance in the mid-plane, $r_0$ is some characteristic radius, $z$ is the distance perpendicular to the mid-plane,
and $h(r)$ is the scaleheight, given by
\begin{equation}
h(r) = h_0 \left(\frac{r}{r_0}\right)^{\beta}
\end{equation}
with parameters of $\alpha=2.125$ and $\beta=1.125$, giving a radial
dependence of the surface density of $\Sigma(r) \propto
r^{-1.0}$. Once the temperature (we assume that the disk is in local
thermodynamic equilibrium passively heated by the star, and gas and dust are thermally coupled) and
dust sublimation (see next paragraph) structures have converged using
the Lucy algorithm, the vertical disk structure is modified via the
equation of hydrostatic equilibrium following a similar algorithm to
that detailed by \citet{walker2004}.  A self-consistent calculation for dust sublimation  and disk temperature followed by a hydrostatic
equilibrium calculation is repeated until the disk density
structure has converged. Convergence is typically achieved in four
iterations. Images and SEDs are subsequently calculated using a
separate Monte Carlo algorithm based on the dust emissivities and
opacities \citep{harries2000}.

The shape of the dust evaporation front is computed following \citet{Tannirk2007}. The dust content is first reduced to a very low value in the computational grid for the circumstellar disk, to make each of the grid cells optically thin. Stellar photons then propagate through the disk
and the temperature of grid cells is determined. Dust is added to
cells that are cooler  (see eqn 3  for sublimation temperature parameterization) than the sublimation temperature in small steps of $\tau$.
 The step size is a $\tau$ of $10^{-3}$ (computed at 5500\AA) for the first five dust growth steps. The step size is then increased
logarithmically, first to $10^{-2}$, then to $10^{-1}$ and so on until
a 100:1 gas to dust ratio is reached in each grid cell.  The
grid cell temperatures are recomputed after every dust growth step and the process is repeated until the shape of the dust sublimation region converges.

We use Kurucz \citep{Kurucz1970} stellar atmosphere models as input spectra for the stars. We consider   a mixture of 0.1,1.3 and 50 micron warm silicates \citep{ossen1992} and power-law opacity mm grains \citep{Man97, Natta2004} to model the opacity in the disk. The micron and sub-micron grain mixture is based on work by \citet{Meeus2001} and \citet{boekel2005}. 
To keep the analysis simple, the grain species are assumed to be well mixed with gas following a delta function size distribution. Dust scattering is not included in the models. Scattering does not seem to have significant impact  on disk structure and interpretation of infrared interferometry for HAe stars \citep{dullemond2003, pinte2008}.

\begin{figure}[h]
\begin{center}
{
\includegraphics[angle=0,width=7.in]{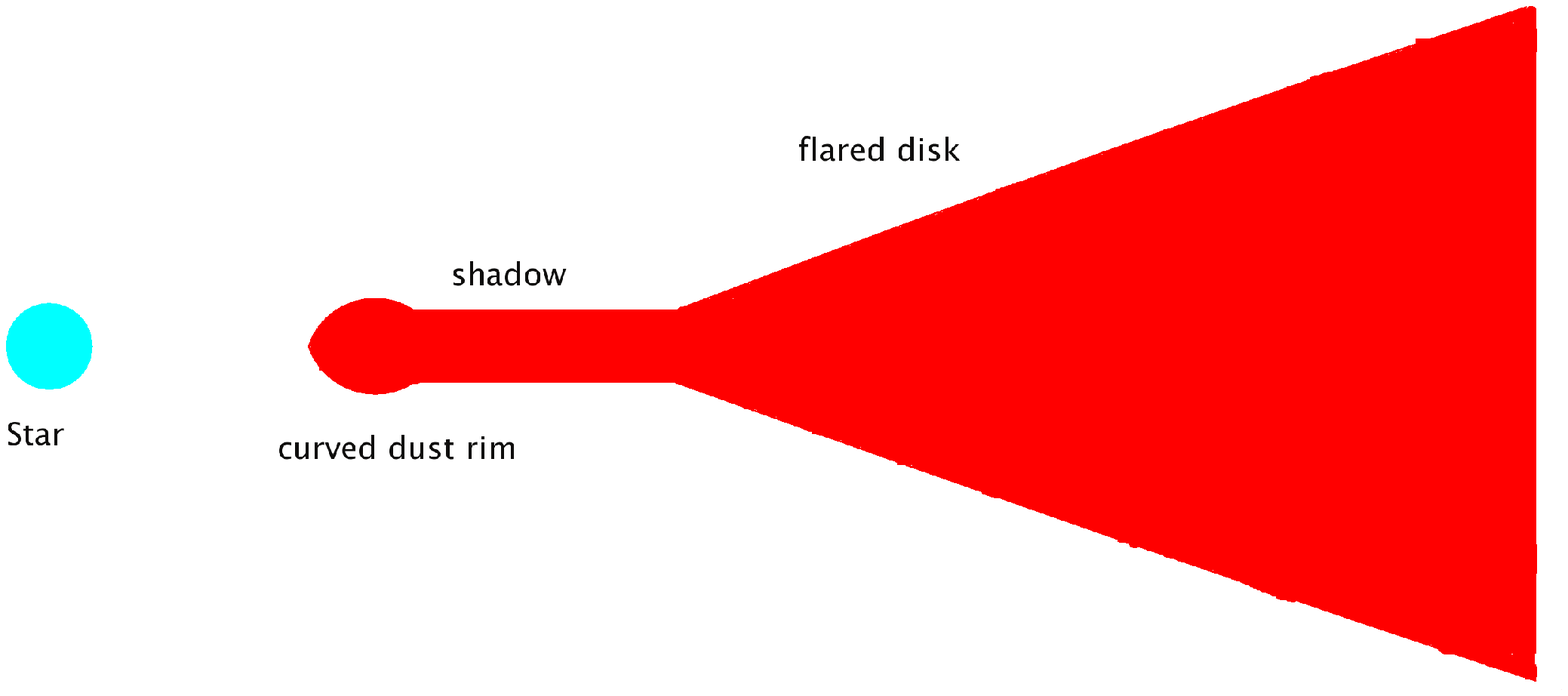}
\hphantom{.....}
\includegraphics[angle=0,width=7.in]{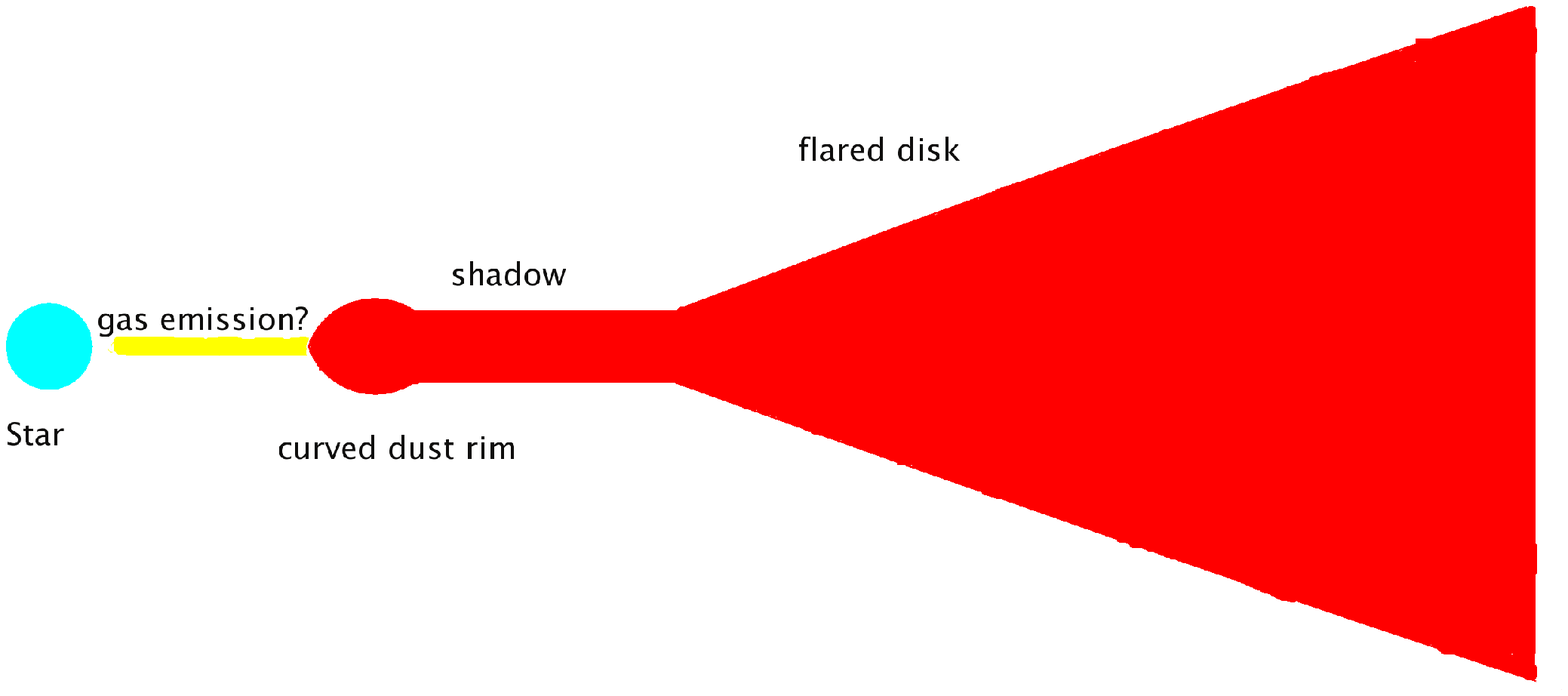}
}

\caption{Schematic of disk models. a)Top panel. Flared disk with a curved inner rim. b) Bottom panel. An additional ``smooth'' emission component (presumably gas) has been added inside the dust destruction radius to explain MWC275 and AB~Aur NIR photometry and interferometry. \newline Note: The models are not to scale.
}
\label{schematic}
\end{center}
\end{figure}

During the course of our modeling and as outlined in \citet{Tannirk2008}, we realized that an additional emission  component (Fig \ref{schematic}),  which we argue to be gas, is needed inside the dust destruction radius to explain  the NIR SED and interferometry  of MWC275 and AB~Aur (see section \ref{MWC275_SED} for discussion on gas opacity). This additional component  is not treated self consistently in our modeling and is added after the dust-opacity-dominated circumstellar-disk model converges in structure. Calculations by \citet{muzerolle2004} have shown that for parameters suitable to MWC275 and AB~Aur, gas does not significantly alter the structure of the dust rim, justifying our simple treatment for the NIR emission geometry. 

In sections \ref{NIR_model_275} and \ref{NIR_model_ABAur},  we demonstrate  that the NIR emitting region in MWC275 and AB~Aur has a radius of $\sim$0.2AU. Furthermore, long-baseline interferometry beyond the first visibility minimum constrains the gas and dust emission to be on similar length scales (Fig~\ref{MWC275_size}). The two statements together imply that in the absence of  shielding of the evaporation front by gas,  the mid-plane sublimation temperature in the dust rim is   $\sim$1850K (see section \ref{NIR_model_275}).

\subsection{Comprehensive models for SED and Interferometry}
To fit the SED and visibilities of MWC275 and AB~Aur we adopted the following algorithm: First, we computed models for the dust evaporation front as described in \S\ref{sublimation_model}. The inner edge of the dust disk is assumed to be dominated by grains larger than  1 micron \citep{lkha2001, isella2006} and the evaporation front shape is set by the density dependence of dust sublimation temperatures \citep{isella2005, Tannirk2007}. The K-band visibilities are computed for the model and compared with data. The normalization of the dust evaporation law  is then adjusted so that the model visibilities fit the visibility data before the first visibility minimum. These models fail to fit the visibility beyond the minimum and do not have sufficient emission to explain the observed NIR SED. Therefore an additional  emission component has been added inside the dust sublimation radius to reconcile the model with the visibility data and NIR SED. 

With the inner-rim parameters fixed, we next  focus on MIR and the mm disk. Millimeter interferometry results from the literature are  used to fix disk masses and sizes. The majority of the dust mass is placed in mm sized grains with a power-law opacity function \citep{Natta2004}. A small fraction ($\sim$10\%) of the dust mass is in micron and sub-micron (small) grains with their relative mass fractions based on literature results \citep{Meeus2001, boekel2005}. The physical extent of  small grains is constrained with MIR imaging and interferometry.  The model is then allowed to run to convergence. The model SED is compared with MIR and far-infrared data, the  mass of the small grain population is then adjusted and models are recomputed until a  good fit to the MIR and far-infrared SED is obtained.  

The MIR visibilities are computed for the SED-converged model and compared with the data and the spatial distribution of the small grain component is adjusted until model visibilities match with data. The only free parameters in our models are the absolute masses of the small grains, mass of the  50$\mu$m silicate grain  and their spatial distribution. Each of the models is computationally expensive. To achieve fast convergence,  the parameter space was varied by hand, until a good fit was found for the observable quantities. 

\section{MWC275: Analysis}
\label{MWC275_model}
MWC275 is a Herbig Ae star (refer to Table \ref{photometry_a}  for basic properties and photometry) with a total luminosity of 36L$_{\odot}$.  The large stellar luminosity, coupled with the fact that the mass accretion rate is $\le$ 10$^{-7}$M$_{\odot}$/year \citep{lopez2006} allows us to ignore accretion heating and model the MWC275 circumstellar disk as a passive disk, reprocessing stellar radiation \citep{cg97, dullemond2001}.  For our models, we choose  the MWC275 disk mass to be between 0.05-0.1 M$_{\odot}$ \citep{Natta2004} and a  surface density profile that varies radially as $r^{-1}$ \citep{isella2007}.  The disk outer edge is truncated at 200AU and  bulk ($\sim$80\%) of the dust mass is assumed to reside in mm grains having  an opacity with a wavelength dependence of $\lambda{^{-1}}$ at long wavelengths. Here, we describe in detail our modeling results for the NIR and MIR morphology of MWC275.
\subsection{The Thermal NIR Disk}
\subsubsection{Visibilities}
\label{NIR_model_275}
Like many other Herbig Ae stars, MWC275 shows a strong NIR excess over stellar photospheric values \citep{Hillen92}. This excess has been traditionally interpreted in terms of the dust disk being truncated by sublimation and forming a `rim'. The rim intercepts stellar photons, re-radiatiing predominantly in the NIR \citep{dullemond2001, isella2005, Tannirk2007}.  However, in T08 we had conclusively shown that models in which all of the NIR excess arises from dust rims alone cannot explain the CHARA interferometry data on MWC275. Our arguments in T08 were necessarily brief. We present a more elaborate analysis in this section.   

\begin{figure}[h]
\begin{center}
{
\includegraphics[angle=90,width=7.5in]{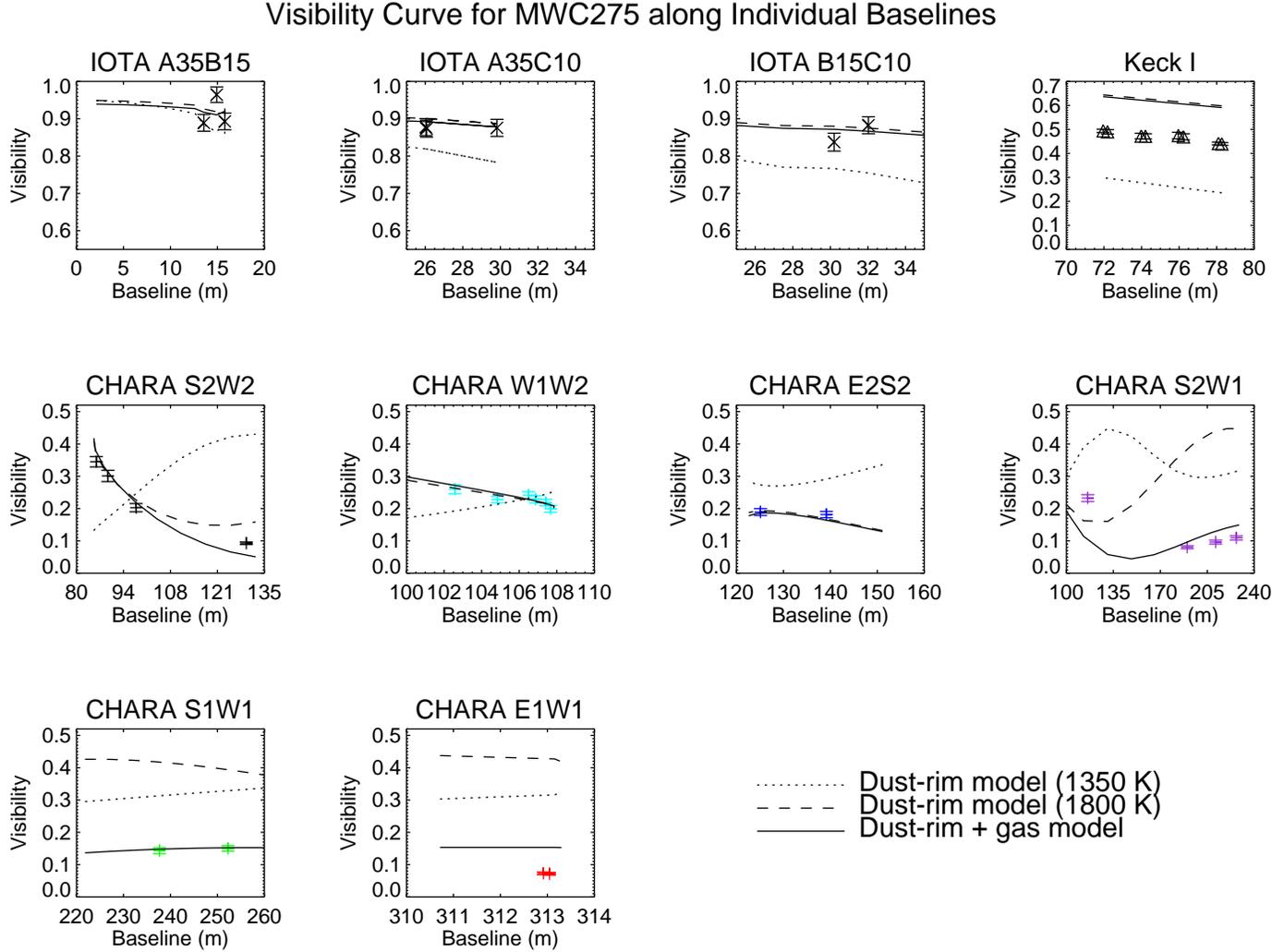}
}

\caption{MWC 275 visibility data and model curves. The quoted model temperatures are at the base of the dust rim.  The NIR size deduced from the Keck Interferometer data (triangles) is $\sim$20\% larger than the size obtained with the CHARA data. This variability of MWC275 is discussed in \S\ref{MWC275_SED}.
}
\label{visibility_total}
\end{center}
\end{figure}

MWC275 observations allow us to clearly detect the asymmetry of the MWC275 disk (see Appendix \ref{appendix2}), as having inclination=48$^{o}{\pm}$2$^{o}$, PA=136$^{o}{\pm}$2$^{o}$, consistent with the inclination of 51$^{+11}_{-9}$ degrees,  PA of 139$^{o}{\pm}$15$^{o}$ determined in \citet{Wassel} and inclination of 46$^o\pm$ 4$^o$,  PA of 128$^o\pm$4$^o$ determined in \citet{isella2007}. The complete visibility data along each of the baselines are presented in Fig \ref{visibility_total}. Following T08, we show the data in a concise manner in  Fig \ref{MWC275_Vis_CHARA} using the notion of an  ``effective baseline'' - $$B_{eff} = B_{projected} \sqrt{{\rm cos}^2(\theta) + {\rm cos}^2(\phi){\rm sin}^2(\theta)}  $$ where $\theta$ is the angle between the uv vector for the observation and the major axis of the disk and $\phi$ is the inclination of the disk. Under the flat disk assumption, the effective baseline  correctly accounts for the change in resolution due to the disk inclination and PA (the geometry of  thick disks is represented only approximately with  optical depth effects and 3-D geometry of thick disks not being taken into account), allowing us to plot the visibility measurements as a function of one coordinate, simplifying presentation and analysis.

We attempt at fitting the  visibilities with a curved inner-rim model (the ``standard'' model) where the rim curvature (variation in cylindrical radius between rim midplane and the atmosphere) is set by the density dependence of dust sublimation temperatures, taken from  \citet{pollack}.  In this model, silicate grains sublimate at a higher temperature compared to other grains and hence fix the rim location. The rim is assumed to be composed of 1.3$\mu$m grains, as larger grains do not  affect the rim shape and location significantly \citep{isella2005}, at the same time  making  numerical convergence slower due to strong back-warming effects \citep{isella2005, Tannirk2007}.  For silicate dust, the evaporation temperature $T_{\rm evp}$ can be parameterized as
 \begin{equation}T_{\rm evp} = G\left[\frac{\rho_{\rm gas}(r,z)}{\mbox{1g }\mbox{cm}^{-3}}\right]^\gamma\end{equation} where $G=2000\K$, $\gamma = 1.95\times10^{-2}$ and $\rho_{\rm gas}$ is the
 density of gas in g\,cm$^{-3}$ (see IN05 eq. [16]). This parameterization, derived from a fit to sublimation temperatures recorded in the laboratory \citep{pollack}, produces a dust rim with an inner edge at 0.36 AU (Fig \ref{rim_atm}a). The rim radius is too large to fit even the relatively short baseline visibility data  from IOTA  (Figs \ref{visibility_total} and \ref{MWC275_Vis_CHARA}). In order to fit the data before the first visibility minimum, we had to increase the T$_{\rm evp}$ normalization-G by $\sim$30\% from 2000K to 2600K. This increases the sublimation temperature at the base of the rim from $\sim$1350K to $\sim$1800K. Fig \ref{rim_atm}b shows the synthetic K-band image for the rim with the increased normalization. The dashed line in Fig \ref{visibility_total} traces the visibility for this model and provides a good fit to the short baseline ($<$ 100m) data.    
 
 However, as seen in Figs \ref{visibility_total} and \ref{MWC275_Vis_CHARA}, rim models which are sharply truncated due to dust sublimation and produce all of the NIR excess, fail to fit observations beyond the first visibility minimum. These models display bounces in visibility at long baselines (not seen in the data)  because of the presence of sharp ring-like features with high spatial frequency components in the corresponding images, even for the smoothest rims physically plausible. In T08, we showed that the presence of a smooth emission component inside the dust destruction radius (Fig \ref{rim_atm}c)  providing 56\% of the total K band emission helps fit the data well (solid line in Figs  \ref{visibility_total} and \ref{MWC275_Vis_CHARA}). The NIR visibility data cannot constrain the surface brightness profile of  the smooth emission component (we have adopted a constant surface brightness profile - a Uniform Disk for simplicity), but can constrain the size scale of the emission fairly robustly. Fig \ref{MWC275_size}  shows a series of visibility curves where the smooth emission component is fixed to be 56\% of the total emission and the radius of the  Uniform Disk component is decreased by 15\% successively from the  initial radius of 0.23AU radius. The model image is then rescaled to maintain a good visibility fit at baselines shorter than 100m. It can be seen in Fig \ref{MWC275_size} that for Uniform Disk (UD) radii smaller than  0.19AU, the model visibilities begin to deviate significantly from the observations. Thus, the CHARA data constrains the smooth emission component to be on the same length scale as the dust sublimation rim filling the region between the disk and the central star.

\subsubsection{SED}
\label{MWC275_SED}
Fig \ref{MWC275_NIR_SED} shows the NIR SED for MWC275. Besides failing to explain the NIR interferometry, the standard model also fails at producing sufficient NIR emission to explain the MWC275 SED even in its `low' state. In T08, we had shown that binarity and source variability cannot account for the discrepancy between the standard model and data. We had argued that the presence of smooth emission inside  the dust destruction radius can help explain the NIR visibility and account for the ``missing'' NIR flux in standard models.  Opacity candidates for the smooth emission component are: (1) a dusty halo around the stars \citep{vinko2006a} and (2) gas inside the evaporation front. 
However, to fit the visibility data, the halo emission would have to be closer to the star than the dust destruction radius in the disk. This would  require even higher dust-sublimation temperatures than the $\sim$ 1850K we are adopting.

The most plausible physical mechanism for the smooth emission  is  hot gas.  The required emission  levels to explain the long-baseline K-band visibility data can be obtained with optically thin gas ($\tau\sim$ 0.15) with a temperature range of 2000K-3000K \citep{muzerolle2004, eisner2007a, eisner2007b}. Assuming that the gas has sufficient opacity to produce the difference in flux between the standard model and  the observed photometry, we can place limits on the wavelength dependence of gas opacity. Fig \ref{opacity_curve}  plots limits on the gas opacity (normalized at K band) such that flux from the gas component + the standard model falls within 10\% of the observed photometry (we have assumed that gas does not significantly alter the geometry of the dust rim). In the next 2 paragraphs we compare theoretical gas-opacity curves with our empirically derived opacity from SED.

Fig. \ref{opacity_curve}  shows the wavelength dependence of molecular  \citep{Ferguson2005, zhu2007} and free-free+free-bound (henceforth FF-BF) opacity, both good candidates for the gas emission (refer T08). At 5000K, FF-BF opacity \citep{Ferguson2005} agrees well with the derived  opacity at long wavelengths but overshoots limits shortward of 2$\mu$m. At temperatures greater than 8000K ,  FF-BF opacities rise quickly with wavelength producing excessive mid-infrared light. 

Theoretical molecular opacities compare fairly with empirical derivation  between 1$\mu$m and 4$\mu$m.  Beyond 4$\mu$m , theoretical molecular opacities rise rapidly with wavelength. However, the observed SED can be matched with  models  only if the  gas opacity  is flat between 4$\mu$m and 9$\mu$m. Also at 2000K and 2500K, strong opacity bands of CO and water vapor are present at 2.5$\mu$m and 5$\mu$m respectively, which have not been observed in MWC275. This suggests that if molecules were contributing to the bulk of NIR opacity, then some of the species providing opacity between 4-8 $\mu$m in \cite{Ferguson2005} and \cite{zhu2007} are being destroyed in the vicinity of Herbig Ae star MWC275.  We note that FF-BF opacities seem to better fit the empirically derived values than molecular opacity.  

\citet{sitko2007} have obtained fairly dense time coverage on the NIR and MIR SED of MWC275. The NIR SED shows variability at the 20\% level. During the same period, the flux in the visible shows no detectable change, indicating that that stellar luminosity remained fairly constant.  \citet{sitko2007} interpret their observations as variations in the structure of the thermal NIR disk. A variation in the NIR morphology  of MWC275 was also detected in the interferometry.  The NIR disk size deduced from the Keck Interferometer data (April 2003 epoch, Figs \ref{visibility_total} and  \ref{MWC275_Vis_CHARA}) is $\sim$20\% larger than the size obtained with the CHARA data (June 2004-Aug 2006 epochs). The size determined from the S2W1 June2007 data also differs at the $\sim$25\% level  from the size obtained from earlier CHARA epochs.   These variations are poorly understood and could be caused by changes in size/geometry, mass accretion rate and gas content in the inner disk. More evidence for  MWC275 variability was recently reported by \citet{Wisniewski2008}, who found changes in  scattered light intensity  between 1998 and 2003-2004.

\subsection{MIR SED and Emission Morphology}
 \label{MIR_model_275}
 \citet{boekel2005} analyzed the MIR SED (Fig \ref{MWC275_SED_total}) of MWC275 in detail and showed that the SED could be reproduced well with a grain mixture of 1.5$\mu$m and 0.1$\mu$m silicates with mass ratio of 4:1. We use results from   \citet{boekel2005} in fixing the small grain composition in our disk models.   
 
As seen in Fig \ref{MWC275_SED_total}, the MWC275 SED falls sharply between 20$\mu$m and 30$\mu$m. This drop and the 10$\mu$m silicate feature can be simultaneously  reproduced only if   the mass fraction of the small grain dust component relative to gas beyond  7AU is less than 20\% of the mass fraction inside of 7AU.  If the small grain component is allowed to exist beyond 7AU, then the model far-infrared spectrum becomes much stronger than observed SED. Fig \ref{MWC275_SED_total} shows a TORUS model SED that fits the MIR and longer wavelength spectrum of MWC275 well. In this model, 40\% of the 8$\mu$m emission arises from the dust rim, with the rim contribution declining to $\sim$20\% at 13$\mu$m. This model also fits the MIDI-VLTI MIR visibilities \citep{Leinert2004} and reproduces the 0.8$\pm$0.1AU 11$\mu$m FWHM minor-axis size of MWC275 (Fig \ref{MIDI}),  naturally explaining  why MWC275 is unresolved by the Keck Segment Tilting Experiment. The quality of the SED and visibility fit in the 8-15$\mu$m region is only moderate, probably due to the simple dust composition and distribution that we have assumed in the model.  The initial model setup has been  chosen to reproduce MWC275 mm-interferometry. 

Table \ref{MWC275_disk_props} lists disk parameters for the MWC275 model and Fig \ref{dust_fraction} shows the radial distribution of the small grain fractions. The mid-plane temperature profile and  the ``flaring'' geometry of the disk surface are shown in Fig \ref{profiles}. The dust rim ``shadows'' \citep{dullemond2001, dullemond2004} the region of the disk between 0.3 and 1AU beyond which the disk begins to flare. The $\tau$=1 surface drops down in scale height steeply after 6.5AU where the small grain fraction reduces sharply.   Our conclusions on dust-grain distribution in the MWC275 disk are consistent with that of \citet{sitko2007}.

\section{AB~Aur: Analysis}
\label{ABAur_model}
AB~Aur is a Herbig Ae star (refer to Table \ref{photometry_b}  for basic properties and photometry) with a total luminosity of 47L$_{\odot}$ \citep{isella2006}.  As in the case of MWC275, AB~Aur's large stellar luminosity  dominates  the circumstellar disk's energy budget \cite[accretion rates $\le$ 10$^{-7}$M$_{\odot}$/year]{lopez2006}.  This  allows us to ignore accretion heating and model the AB~Aur circumstellar disk as a passive disk, reprocessing stellar radiation \citep{cg97, dullemond2001}.  For our models, we choose  the AB~Aur disk mass to be between 0.007-0.013 M$_{\odot}$ \citep{lin2006} and a surface density profile that falls radially as $r^{-1}$ \citep{corder2005}.  The disk outer edge is truncated at 300AU and the bulk ($\sim$80\%) of the dust mass is assumed to reside in mm grains with an opacity that depends on wavelength as $\lambda{^{-1}}$ for long wavelenghts. Here, we describe in detail our modeling results for the NIR and MIR morphology, and SED of  AB~Aur.

\subsection{The Thermal NIR Disk}
\label{NIR_model_ABAur}
We follow the procedure outlined in \S\ref{NIR_model_275} to model the SED and visibilities of AB~Aur. We first attempt at fitting a standard curved dust rim model (Fig \ref{rim_atm_faceon}) to the NIR visibilities. The rim is assumed to be composed of 1.3$\mu$m silicate grains and the dust evaporation temperature law is described by equation (3). This produces rim radii too large to fit baselines shorter than 100m and we had to increase the T$_{evp}$ normalization to 2800K from 2000K. This increases the sublimation temperature at the base of the rim from $\sim$1350K to $\sim$1950K. The dashed line in Fig \ref{ABAur_Vis_CHARA} traces the visibility for this model and provides a good fit to the data at baselines shorter than 100m.

The dust-rim-only  model produces large bounces in visibility beyond 150m and as in the case of MWC275, this bounce is not observed. We have scanned the 150m-300m baseline (Fig \ref{ABAur_Vis_CHARA}) range several times with CHARA and have failed to detect fringes, ruling out  dust-rim only models for the AB~Aur NIR emission. The addition of a Uniform Disk  of emission interior to the dust destruction radius (Fig \ref{rim_atm_faceon}c) helps fit the data well (solid line in Fig \ref{ABAur_Vis_CHARA}). The gas component also helps fit the NIR SED (Fig \ref{ABAur_NIR_SED}). Parameters for the `dust rim + Uniform Disk'  model are listed in  Table \ref{AB_Aur_disk_props}.  

\subsection{MIR SED and Emission Morphology}
 \label{MIR_model_ABAur}
  \citet{liu2007} resolved the AB~Aur disk at 10.3$\mu$m using nulling interferometry and measured a disk is inclination  of 45$^o$- 65$^o$ inconsistent with nearly face on measurements in the mm \citep{corder2005} and the NIR \citep{rmg2001, eisner2004}. \citet{liu2007} interpreted their result in terms  of the AB~Aur circumstellar environment  being more complicated than a disk. Since AB~Aur is well resolved by the Keck Segment Tilting Experiment (Fig \ref{ABAur_LWS}), a disk inclination of 45$^o$- 65$^o$ would have produced observable size difference between the major and minor axis of the disk.  We do not find evidence for this size variation in  our Segment Tilting data, and hence support a face on model for the mid-infrared disk around AB~Aur consistent with the NIR and mm results.
  
  The MIR spectrum of AB~Aur in the 10.7$\mu$m to 20$\mu$m range can be modeled well with  a dust grain mixture of 1.3$\mu$m and 0.1$\mu$m silicates with equal mass fractions \citep{boekel2005}. In addition to the micron and sub-micron silicates, we include a 50$\mu$m silicate component to model the relatively flat spectrum of AB~Aur between 35$\mu$m and 80$\mu$m .

Fig \ref{ABAur_SED_total} shows a TORUS model SED that fits the MIR and longer wavelength spectrum of AB~Aur well . In this model, $\sim$40\% of the 8$\mu$m emission arises from the dust rim, with the rim contribution declining to $\sim$10\% at 13$\mu$m. This model also fits the Keck Segment Tilting data visibilities reproducing the 10.5$\pm$0.7AU 10$\mu$m FWHM size of AB~Aur (Fig \ref{ABAur_LWS}).  By initial design, the model  fits AB~Aur mm-interferometry and SED. 

Table \ref{AB_Aur_disk_props} lists disk parameters for the AB~Aur model and Fig \ref{dust_fraction} shows the radial distribution of the small grain fractions. The mid-plane temperature profile and the $\tau$=1 surface at 5500\AA~ are shown in Fig \ref{profiles}. The inner rim shadows the disk between 0.3AU  and 1AU, beyond which the disk surface takes a flared geometry.

\section{Discussion}
\label{discuss}
The simultaneous modeling of the infrared and millimeter SED and interferometry of MWC275 and AB~Aur allows us to address several important issues regarding  the structure of their circumstellar disks.  To maintain clarity in our discussion we divide the disk into two regions (i) thermal NIR region ($<$ 0.3 AU) (ii) outer disk (between 0.3AU and the disk outer edge).

\subsection{The Thermal NIR Disk}
 Detailed modeling (\S\ref{NIR_model_275} and \S\ref{NIR_model_ABAur}) of the inner disk shows that models where bulk of the NIR emission arises in a dust rim   truncated by sublimation  fail to fit the long-baseline interferometry data  and under-estimate the NIR emission by a factor of 2 relative to observations.  As mentioned in T08 and demonstrated  in detail in this work, the presence of a gas emission component inside the dust destruction radius can solve the interferometry and SED problem simultaneously. 

This however opens up a number of new questions,  namely (i) What is the geometry of the gas dust transition region? To date there has been no calculation of transition region structure that treats both gas and dust simultaneously in a self consistent manner.  (ii) What are the relative contributions of accretion and stellar radiation to heating the gas? We have shown that an ad-hoc addition of an NIR emission component inside the dust destruction radius helps explain the data, but the current modeling does not shed any light on the energy budget question. (iii) What are the gas species that provide the NIR opacity?  Is the opacity molecular in nature or is it from free-free and free-bound processes? If a significant portion of the NIR emission were indeed arising from molecular gas,  then Fig \ref{opacity_curve} shows  that theoretical gas opacities depend much more sensitively on wavelength between 4 and 10$\mu$m than what is observed. This suggests that some of the molecules providing the model opacities might be getting destroyed by the stellar UV radiation field. 

In the course of modeling the MWC275 and AB~Aur disks, we realized (\S\ref{NIR_model_275}) that the observed K-band sizes could be reproduced only if the dust sublimation temperature at the base of the dust rims were increased to $\sim$1850K from the experimentally measured silicate evaporation temperatures of $\sim$1400K \citep{pollack}.  A simple treatment of the gas-dust transition region by \citet{muzerolle2004} suggests that gas is not effective in modifying rim geometry.  In the absence of shielding by gas, the  large dust sublimation temperatures indicate that  the grains in the inner disks of young stars are significantly more refractory and/or optically transparent than has been assumed in the literature. There is also the possibility that the gas gets optically thick along the mid-plane, shielding the dust from direct stellar radiation  and allowing the dust rim to exist closer to the star \citep{monnier2005, isella2006}.   Future, high resolution NIR spectroscopic studies  of MWC275 and AB~Aur,  combined with self consistent  models of the gas  density and temperature structure, will help address many of the questions raised here.

MWC275 and AB~Aur require gas emission to explain the their SED interferometry. In contrast,  past modeling work by \cite{isella2006} seems to suggest that dust rims alone are probably sufficient to explain the NIR data on the young stars V1295 Aql (A2 IVe) and CQ Tau (F2 IVe). A larger sample of young  stars will therefore have to be observed with milli-arcsecond interferometry to establish and understand trends between spectral type, stellar mass, accretion rates and the contribution of gas emission to NIR SED.  

A new and exciting observational domain will be opened  with the commissioning of the fringe tracker \citep{Berger2006} for CHARA-MIRC \citep{monnier2007} in the summer of 2008. This will sufficiently   improve CHARA-MIRC sensitivities to combine light from 3 or more telescopes, allowing the first milli-arcsecond non-parametric imaging of MWC275 and AB~Aur in the NIR. The snapshot multiple-baseline coverage will provide us a powerful tool in understanding the  infrared time variability of YSO disks.

\subsection{The Outer Disk}
 Our models for the MWC275 and AB~Aur MIR interferometry and SED suggest that the outer disks of these systems are at different evolutionary stages. MWC275 10 micron size and MIR SED can only be reproduced if the disk is depleted in  micron and sub-micron sized grains beyond $\sim$7AU (\S\ref{MIR_model_275}). This meshes well with the fact that the observed 10.7$\mu$m size of MWC275 is $\sim$3 times  smaller than AB~Aur. The depletion of small grains beyond 7AU in the disk atmosphere  indicates that the dust particles  in MWC275 have undergone significant settling. However, the presence of the 10$\mu$m silicate feature in MWC275 implies that there is some process (like planetesimal collisions) that maintains the supply of micron sized grains in the inner regions of the disk. 

Our models predict that the inner dust rim shadows \citep{dullemond2004}  the region of the disk between 0.3AU and 1AU. The structure and size of the shadow depends sensitively on the composition of grains in the circumstellar disk \citep{Tannirk2007} and hence is an important probe of dust physics. The presence of the shadow has not been observationally confirmed yet in any YSO system, although some indirect evidence  has been found in VV Ser \citep{ponto2007}.

\section{Conclusions} 
\label{conclusion}
We have presented the first set of comprehensive disk models for the SED and interferometry of Herbig Ae stars MWC275 and AB~Aur. We have shown that  `standard'  models  for the dust evaporation front   where the bulk of the near-infrared emission arises from a dust wall, fail to explain the near-infrared spectral energy distribution and interferometry. Standard  models produce large bounces in visibility at high spatial frequency, which is not observed in the data.  We have conclusively demonstrated that the presence of an additional smooth emission component (presumably hot gas) inside the dust destruction radius and on a similar size scale to the dust rim  can ameliorate the situation. In the absence of shielding of star light by gas, we have established that dust grains in the gas-dust transition region will have to be highly refractory, sublimating at ~1850K. The small mid-infrared size of MWC275 relative to AB~Aur, shows that the dust grains in the outer disk MWC275 are  significantly more evolved/settled than the  grains  in the AB~Aur disk. We suggest that dynamical processes (like planetesimal collisions) that maintain the  population of micron-sized grains producing the 10$\mu$m feature in the spectrum, are operational only in the inner ~7AU of MWC275. However, in AB-Aur the small-dust producing mechanisms exist at least out to ~20 AU and maybe even beyond.

\acknowledgements
AT acknowledges contributions from Nuria Calvet, Michael Busha, Marlin Whitaker and Steve Golden. Research at the CHARA array is supported by  the National Science Foundation through grants AST 06-06958 and AST 03-52723 and by the Georgia Sate University through the offices of the Dean of the College of Arts and Sciences and the Vice President for Research. This project was partially supported by NASA grant 050283 and NSF grant AST 03-52723 .  This publication makes use of NASA's Astrophysics Data System Abstract Service. CHARA visibility-calibrator sizes were obtained with the fBol module of getCal, a package made available by the Michelson Science Center, California Institute of Technology (http://msc.caltech.edu). Computations were performed on the Legato-Opus Cluster Network at the University of Michigan.

\begin{table}[h]
\begin{center}
\caption{CHARA uv coverage and visibility data for MWC275. The array geometry is illustrated in Fig 1, \citet{ten2005}.\newline
}
\begin{tabular}{cccccc}
\footnotesize
UT-Date        & u(m) & v(m) & Telescope  & Calibrated & Calibrator \\
of Observation &               &               &   pair     &  Visibility & Names\\
\hline
2004July09    & -210.61  & 138.79 &   S1W1 & 0.150$\pm$0.008 & HD164031    \\
              & -200.38  & 127.78 &        & 0.143$\pm$0.009 &   \\
2005July22    &  106.91  & -11.88 &   W1W2 & 0.218$\pm$0.011 & HD164031   \\
              &  103.22  & -18.38 &        & 0.227$\pm$0.009 &   \\
2005July26    &  102.45  &  5.03 &   W1W2 & 0.260$\pm$0.014  & HD164031  \\
              &  106.45  & -1.61 &        & 0.241$\pm$0.011  &  \\
              &  107.18  & -10.26 &        & 0.201$\pm$0.011 &   \\
              &  105.94  & -13.95 &        & 0.232$\pm$0.011 &   \\
2006June22    & -11.99   & 84.94 &   S2W2 & 0.345$\pm$0.016  & HD164031    \\
              & -24.60  & 85.73 &        & 0.301$\pm$0.017   &  \\
              & -42.30  & 87.77 &        & 0.203$\pm$0.013   &  \\
2006June23    & -301.23 &-85.20 &   E1W1 & 0.0715$\pm$0.0043 & HD164031   \\
              & -302.93 & -78.34 &        & 0.0730$\pm$0.0044&    \\
2006June23    & -84.05  & 98.77 &   S2W2 & 0.0925$\pm$0.0041 & HD164031  \\
2006Aug23     &  60.15  & 125.48 &   E2S2 & 0.181$\pm$0.010  & HD164031, HD166295  \\
              &  28.16  & 121.90 &        & 0.189$\pm$0.011  & \\
2007June17    & -94.21 & 66.98  &   S2W1 & 0.232$\pm$0.013   & HD164031, HD156365 \\
              & -166.86 & 90.88  &        & 0.080$\pm$0.005  & \\
              & -184.56 & 102.73  &        & 0.096$\pm$0.006 &  \\
              & -195.84 & 114.04  &        & 0.110$\pm$0.007 &  \\
\hline
\end{tabular}
\label{uv_table1}
\end{center}

\end{table}

\begin{table}[h]
\begin{center}
\caption{CHARA uv coverage and visibility data for AB~Aur.
}
\begin{tabular}{cccccc}
\footnotesize
UT-Date        & u(m) & v(m) & Telescope & Calibrated  & Calibrator \\
 of Observation&               &               &    pair   &   Visibility & Names\\
\hline
2006Aug23 & 212.04&  237.05& E1S1 & 0.095$\pm$0.005   & HD29645, HD31233    \\
          & 203.79&  251.87&      & 0.120$\pm$0.006   &      \\
          & 197.53&  259.95&      & 0.123$\pm$0.007   &      \\
2006Dec14 & -5.77 & -325.14& E1S1 & 0.115$\pm$0.007   & HD29645, HD31233     \\
2006Dec15 & -93.57&  8.26& E2W2 & 0.188$\pm$0.011     & HD29645, HD31233    \\
\hline
\end{tabular}
\label{uv_table2}
\end{center}

\end{table}

\begin{table}[h]
\begin{center}
\caption{Keck Segment Tilting Experiment   baseline coverage and uv averaged visibility data for MWC275 and AB~Aur.
}
\begin{tabular}{ccccc}
\footnotesize
UT-Date        & Baseline(m) &  Calibrated  & Calibrator \\
 of Observation&                  &   Visibility & Names\\
\hline
&  & {MWC275\hspace{2cm}} & & \\
\hline
2004Sep01 & 3.03 & 0.969$\pm$0.049 & v3879 Sgr \\
'' & 4.72 & 0.944$\pm$0.040 & '' \\
'' & 5.49 & 0.946$\pm$0.036 & '' \\
'' & 7.21 & 0.942$\pm$0.033 & '' \\
'' & 8.43 & 0.963$\pm$0.033 & '' \\
\hline
&  & {AB~Aur\hspace{2cm}} & & \\
\hline
2004Aug30, 31 \&  & 3.03 & 0.870$\pm$0.039 & iota Aur \\
Sep01 & & & \\
'' & 4.72 & 0.823$\pm$0.027 & '' \\
'' & 5.49 & 0.807$\pm$0.033 & '' \\
'' & 7.21 & 0.753$\pm$0.047 & '' \\
'' & 8.43 & 0.708$\pm$0.039 & '' \\
\hline
\end{tabular}
\label{LWS}
\end{center}

\end{table}

\begin{deluxetable}{cc}
\tablecolumns{2}
\tablewidth{0pc}
\tablecaption{Basic stellar properties and photometry for MWC275. \label{photometry_a} }
\tablehead{
\colhead{Property} & \colhead{Value} }
\startdata  
RA   & 17~56~21.29 \\
Dec  & $-$21 57 21.8 \\
Spectral Type  &  A1e$^a$    \\
T$_{\rm eff}$      &        9500K$^a$ \\
Luminosity   &        36 L$_{\odot}$$^a$\\
Distance       &       122pc$^a$\\
Mass              &       2.3M$_{\odot}$$^a$\\
U                 &         -   \\
B                 &         6.98$\pm$.08$^b$  \\
V                 &        6.84$\pm$.06$^b$    \\
R                 &     6.86$\pm$.05$^b$  \\
I                   &      6.71$\pm$.07$^b$   \\
J                  &       6.20$\pm$.08$^b$   \\
H                 &       5.48$\pm$.07$^b$\\
K                  &    4.59$\pm$.08$^b$\\  

\enddata

\tablenotetext{a} {Stellar parameters from \cite{monnier2006}, \cite{Natta2004} and references therein.}
\tablenotetext{b} {Photometry obtained at MDM Observatories (longitude: $-$111.67$^o$, latitude:31.95$^o$) in 2006 June.}

\end{deluxetable}

\begin{deluxetable}{cc}
\tablecolumns{2}
\tablewidth{0pc}
\tablecaption{Basic stellar properties and photometry for AB~Aur. \label{photometry_b}}
\tablehead{
\colhead{Property} & \colhead{Value} }
\startdata  
Spectral Type  &  A0pe$^a$  \\
T$_{\rm eff}$      &        9772K$^a$ \\
Luminosity   &        47 L$_{\odot}$$^a$\\
Distance       &       144pc$^a$\\
Mass              &       2.4M$_{\odot}$$^a$\\
U                 &          7.18$\pm$.08$^b$ \\
B                 &           7.14$\pm$.04$^b$ \\
V                 &         7.01$\pm$.04$^b$ \\
R                 &      6.96$\pm$.05$^b$\\
I                   &        6.70$\pm$.09 $^b$ \\
J                  &         5.99$\pm$.05$^b$  \\
H                 &       5.28$\pm$.05$^b$\\
K                  &      4.37$\pm$.05$^b$\\  
\enddata
\tablenotetext{a} {Stellar parameters from \cite{monnier2006}, \cite{isella2006} and references therein.}
\tablenotetext{b} {Photometry obtained at MDM Observatories in 2005 December.}
\end{deluxetable}

\begin{deluxetable}{ccc}
\tablecolumns{3}
\tablewidth{0pc}
\tablecaption{MWC275 disk properties from the literature \label{MWC275_properties}}
\tablehead{
\colhead{}   & \colhead{Dust Disk}  &  \colhead{} }
\startdata  
Mass  &  & 0.0007M$_{\odot}$$^a$\\
Dust to Gas Ratio & &  0.01   \\
Surface Density Profile & &  $r^{-1}$  $^{a,b}$\\
 Outer Radius  &  &  200 AU $^b$ \\
 Inclination  &  &  48$^o\pm$2$^o$  (this work, c)\\
 Position Angle &  &  136$^o\pm$2$^o$ (this work, c) \\
\cutinhead{Relative Mass Fractions of micron and sub-micron grains in the disk atmosphere$^d$} 

  0.1 $\mu$m silicates & 1.5 $\mu$m silicates & PAH    \\
     0.19$^{+0.009}_{-0.018}$ &    0.8$^{+0.05}_{-0.04}$    &    0.01$^{+0.001}_{-0.001}$   
\enddata

\tablenotetext{a} {\cite{Natta2004}}
 \tablenotetext{b} {\cite{isella2007}}
 \tablenotetext{c} {\cite{Wassel}}
 \tablenotetext{d} {\cite{boekel2005} }

\end{deluxetable}

\clearpage

\begin{deluxetable}{ccc}
\tablecolumns{3}
\tablewidth{0pc}
\tablecaption{AB~Aur disk properties from the literature \label{ABAur_properties}}
\tablehead{
\colhead{}   & \colhead{Dust Disk}  &  \colhead{} }
\startdata  
Mass  &  & 0.0001M$_{\odot}$$^a$\\
Dust to Gas Ratio & &  0.01   \\
Surface Density Profile & &  $r^{-1}$$^b$\\
 Outer Radius  &  &  300 AU$^a$ \\
 Inclination  &  &  21$^o\pm$0.5$^o$$^b$\\
 Position Angle &  &  58.6$^o\pm$0.5$^o$$^b$ \\
\cutinhead{Relative  Mass Fractions of micron and sub-micron grains in the disk atmosphere$^c$} 

  0.1 $\mu$m silicates & 1.5 $\mu$m silicates & PAH    \\
     0.5$^{+0.03}_{-0.03}$ &    0.48$^{+0.03}_{-0.04}$    &    0.02$^{+0.001}_{-0.002}$   
\enddata

\tablenotetext{a} {\cite{lin2006}}   
\tablenotetext{b} {\cite{corder2005}}
\tablenotetext{c}{\cite{boekel2005}}

\end{deluxetable}

\clearpage

\begin{deluxetable}{ccc}
\tablecolumns{3}
\tablewidth{0pc}
\tablecaption{MWC275 model-disk  properties constrained by this work. \label{MWC275_disk_props}}
\tablehead{
\colhead{}   & \colhead{Dust Disk}  &  \colhead{}  }
\startdata  
Inner Radius   &  &   0.22 AU$^a$  \\
K-band flux contribution from dust rim                 &            &  29\% $^a$ \\
Mass fractions of dust components &  & see Fig \ref{dust_fraction} \\
\cutinhead{ \hspace{2.3cm}            NIR Gas Disk}
Surface Brightness Profile            &             & constant  \\
& &                                                    (poorly constrained)  \\                          
Outer Radius                                      &              &  0.22 AU $^a$ \\
K-band flux contribution                  &            &  56\% $^a$ \\              
Temperature                                      &             &  $>$ 1800K  \\
 Vertical Optical Depth                     &             &  0.15   $^a$   \\
Gas-Opacity Profile                          &              & see Fig \ref{opacity_curve} \\
\enddata

\tablenotetext{a} {\cite{Tannirk2008}. The star contributes 10\% of the K-band flux and an extended envelope \citep{monnier2006} contributes 5\%.}
\end{deluxetable}

\begin{deluxetable}{ccc}
\tablecolumns{3}
\tablewidth{0pc}
\tablecaption{AB~Aur model-disk  properties constrained by this work.\label{AB_Aur_disk_props}}
\tablehead{
\colhead{}   &   \colhead{Dust Disk}   & \colhead{}}
\startdata  
Inner Radius & &    0.24 AU $^a$  \\
K-band flux contribution from dust rim                 &            &  22\% $^a$ \\
Mass fractions of dust components & & refer Fig \ref{dust_fraction} \\
\cutinhead{NIR Gas Disk}
Surface Brightness Profile          &             & constant  \\
& &                                                (poorly constrained)  \\                          
Outer Radius                                 &               &  0.24 AU $^a$  \\
K-band flux contribution               &               &  65\%   \\              
Temperature                                   &               &  $>$ 1900K  \\
 Vertical Optical Depth                   &              &  0.14  $^a$    \\
Gas-Opacity Profile                        &                & refer Fig \ref{opacity_curve} \\
\enddata

\tablenotetext{a} {\cite{Tannirk2008}. The star contributes 8\% of the K-band flux and an extended envelope \citep{monnier2006} contributes 5\%.}
\end{deluxetable}

\clearpage

\clearpage

\clearpage

\begin{figure}[h]
\begin{center}
{
\includegraphics[angle=90,width=7.in]{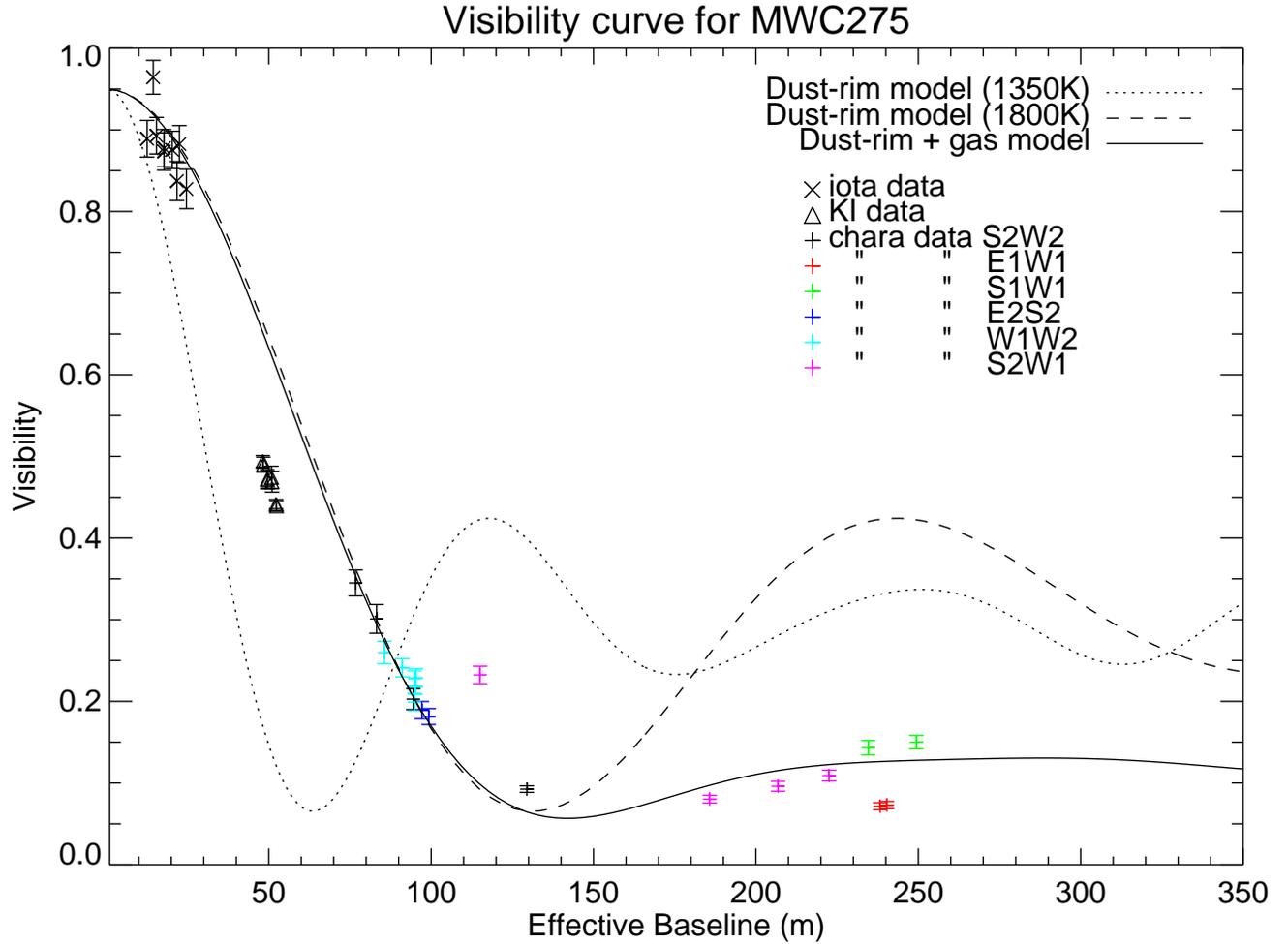}
}
\caption{MWC275 visibility vs `Effective Baseline'. Effective Baselines are  useful in presenting data along multiple uv vectors in a concise manner (under the assumption of axial symmetry). The NIR size deduced from the Keck Interferometer data (triangles) is $\sim$20\% larger than the size obtained with the CHARA data. This variability of MWC275 is discussed in \S\ref{MWC275_SED}
}
\label{MWC275_Vis_CHARA}
\end{center}
\end{figure}

\begin{figure}[h]
\begin{center}
{
\includegraphics[angle=90,width=3.in]{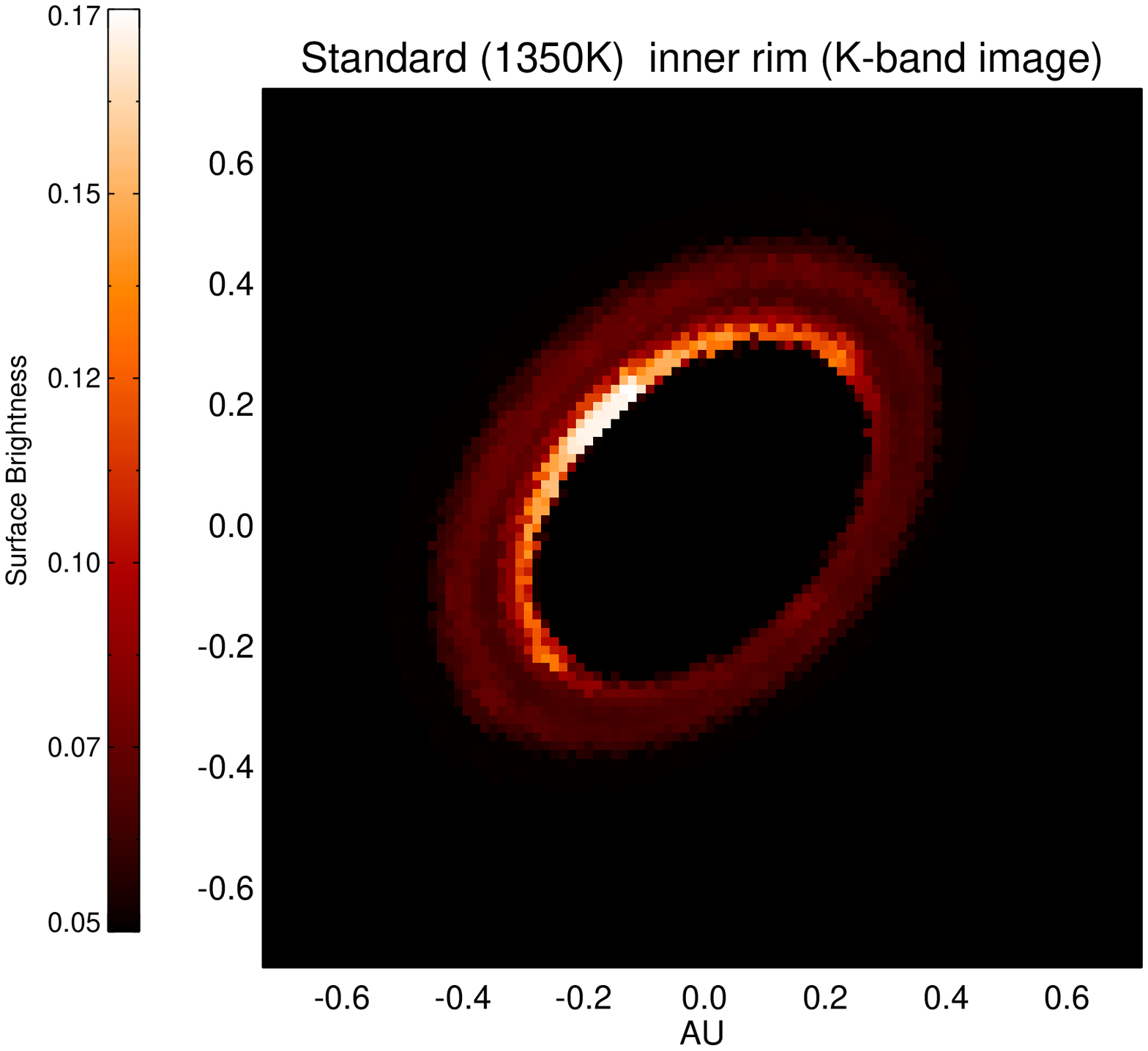}
\hphantom{.....}
\includegraphics[angle=90,width=3.in]{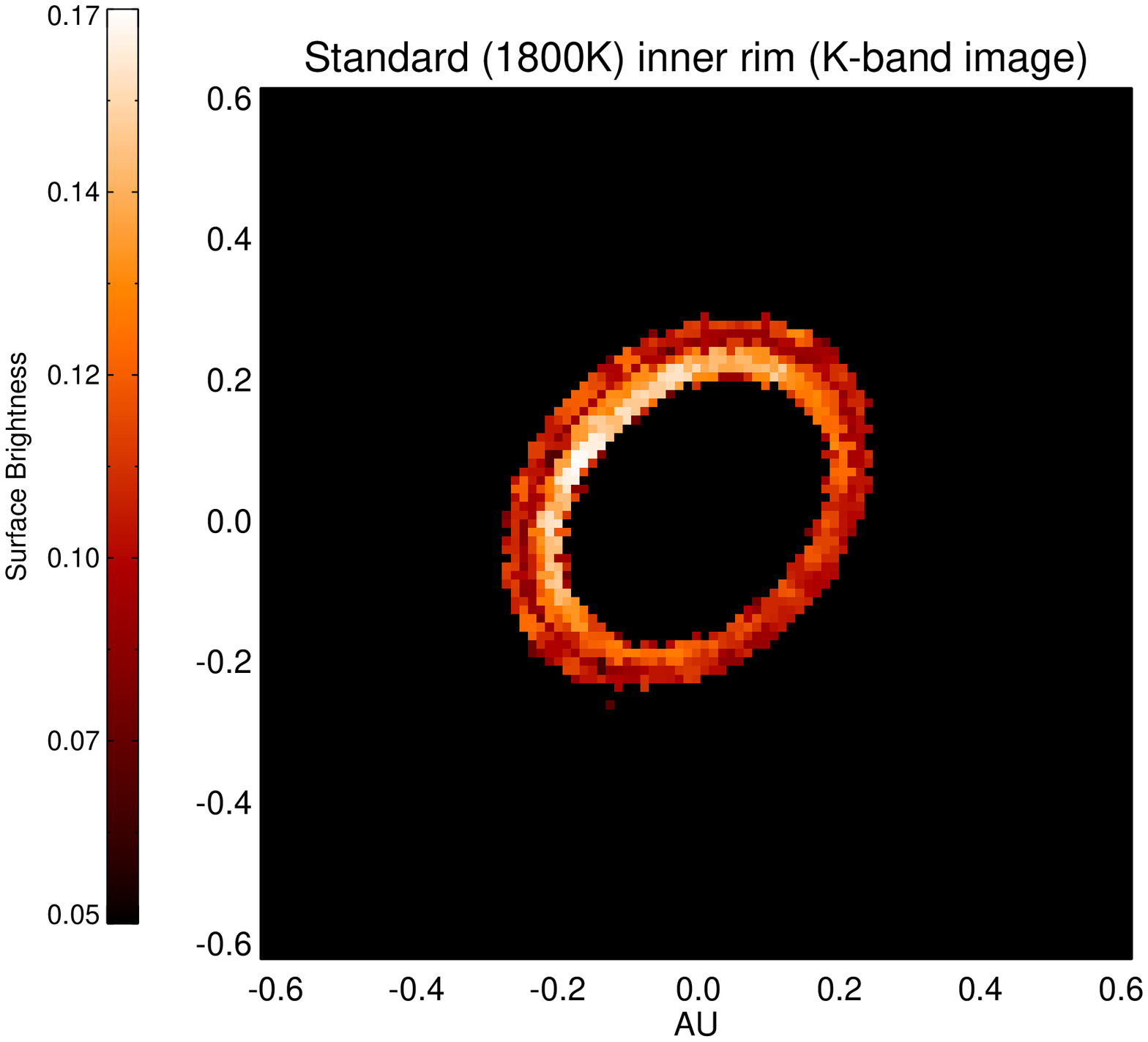}
\hphantom{.....}
\includegraphics[angle=90,width=3.in]{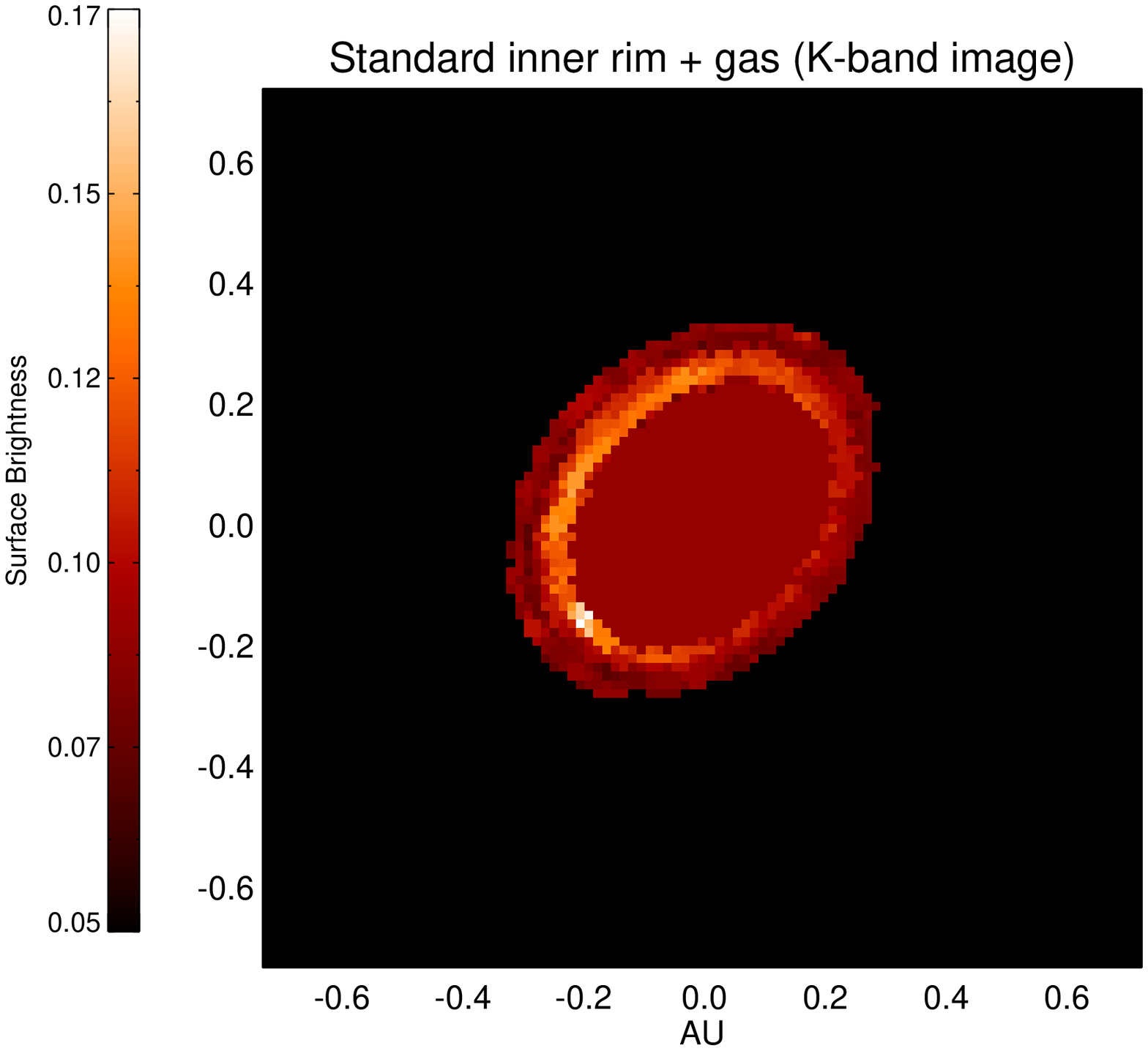}
}

\caption{Inclined-disk models for NIR emission in MWC275.  The disk has an inclination of  48$^o$ and a  PA of 136$^o$ (North is towards the  top and East  is on the left). The sense of the inclination is from \citet{grady1999} a) Top  left panel. A  standard curved dust-rim-only model with rim-base temperature  $\sim$1350K.  b) Top  right  panel. Standard curved dust-rim-only model with  rim-base temperature $\sim$1800K.
c) Bottom panel. Curved dust-rim model with gas emission (modeled as a uniform disk centered on the star)  inside  the dust rim to smooth out the emission profile.
}
\label{rim_atm}
\end{center}
\end{figure}

\begin{figure}[h]
\begin{center}
{
\includegraphics[angle=90,width=5.in]{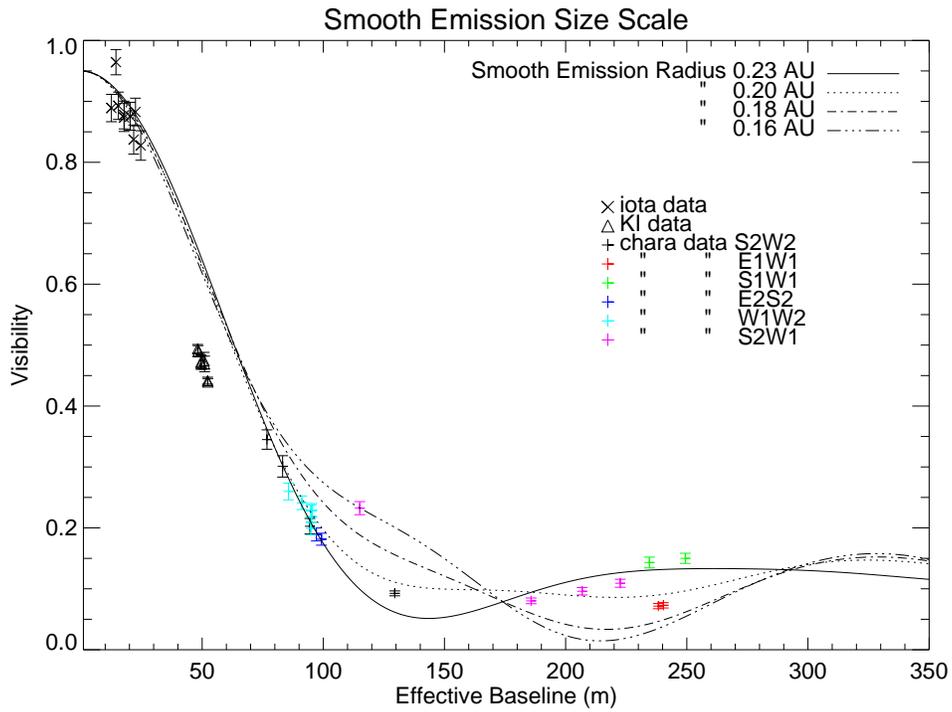}
}
\caption{Constraining the size scale of the smooth emission component interior to the dust destruction radius in MWC275. The model visibilities begin to deviate significantly from the data when the radius of the smooth emission component  becomes smaller than 0.19 AU.  
}
\label{MWC275_size}
\end{center}
\end{figure}

\begin{figure}[h]
\begin{center}
{
\includegraphics[angle=90,width=4.in]{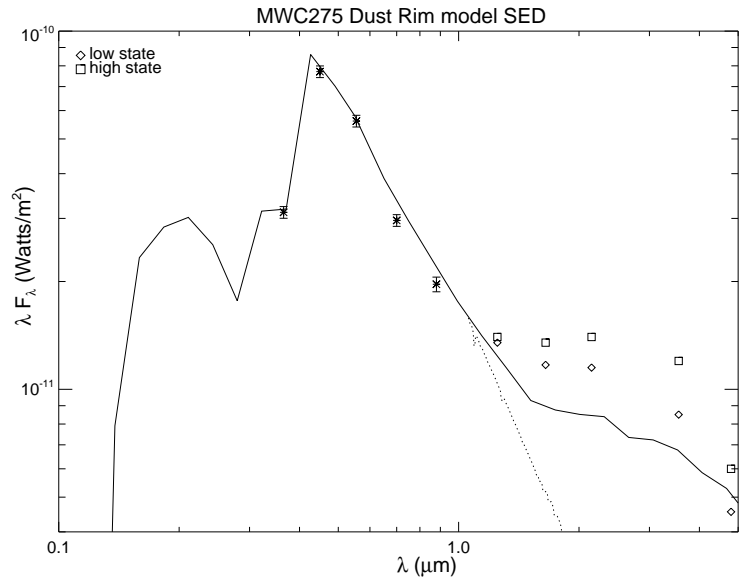}
}
\caption{The NIR SED for MWC275. The `stars' are photometry points from MDM (Appendix Table \ref{photometry_a}). The  `squares' and `diamonds' are high and low state measurements from \citet{sitko2007}. The solid line is the SED produced by the `star + dust-rim only' model in Fig \ref{rim_atm}b.  The dotted line is the SED of the star.
}
\label{MWC275_NIR_SED}
\end{center}
\end{figure}

\begin{figure}[h]
\begin{center}
{
\includegraphics[angle=90,width=4.in]{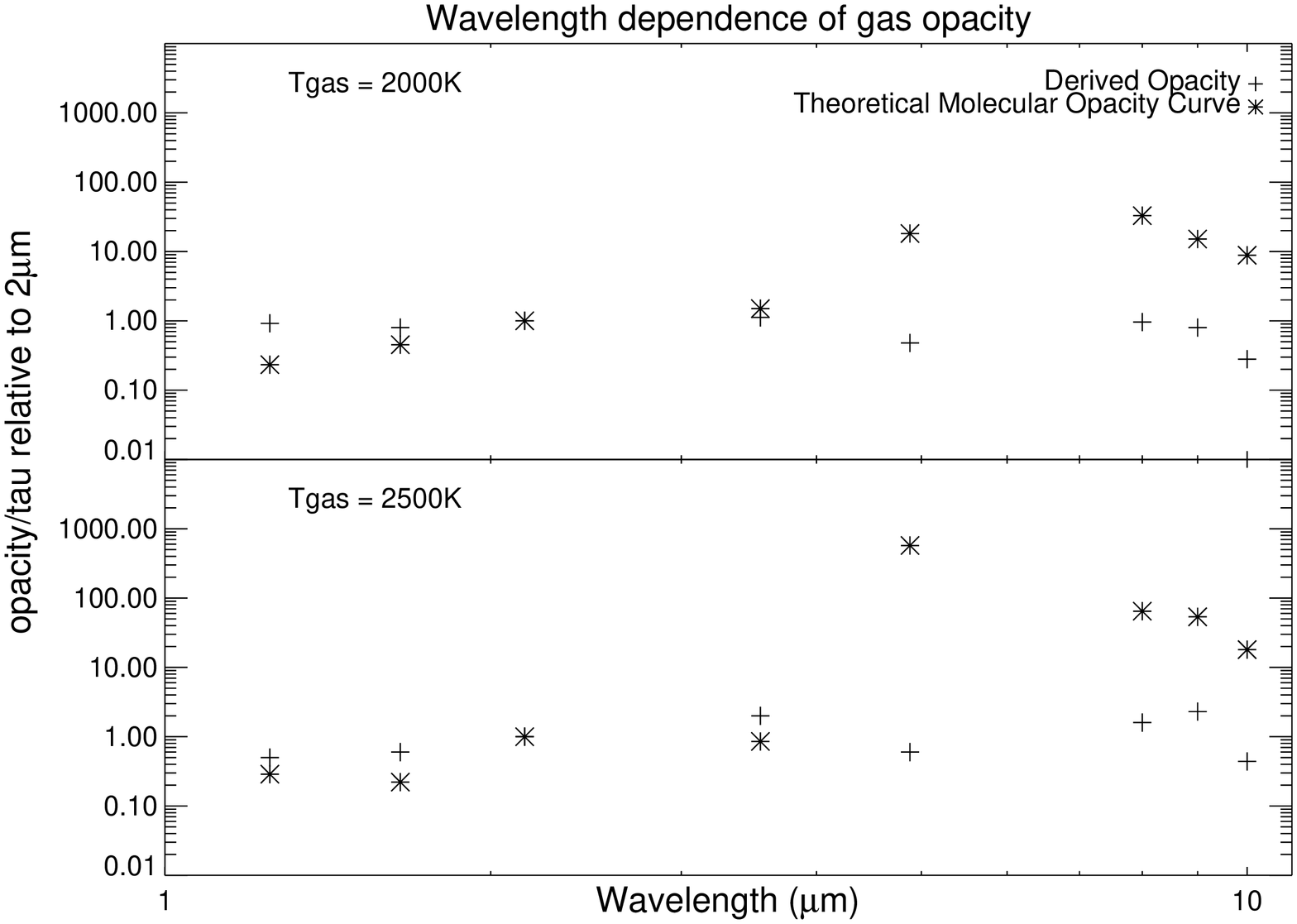}
\hphantom{.....}
\includegraphics[angle=90,width=4.0in]{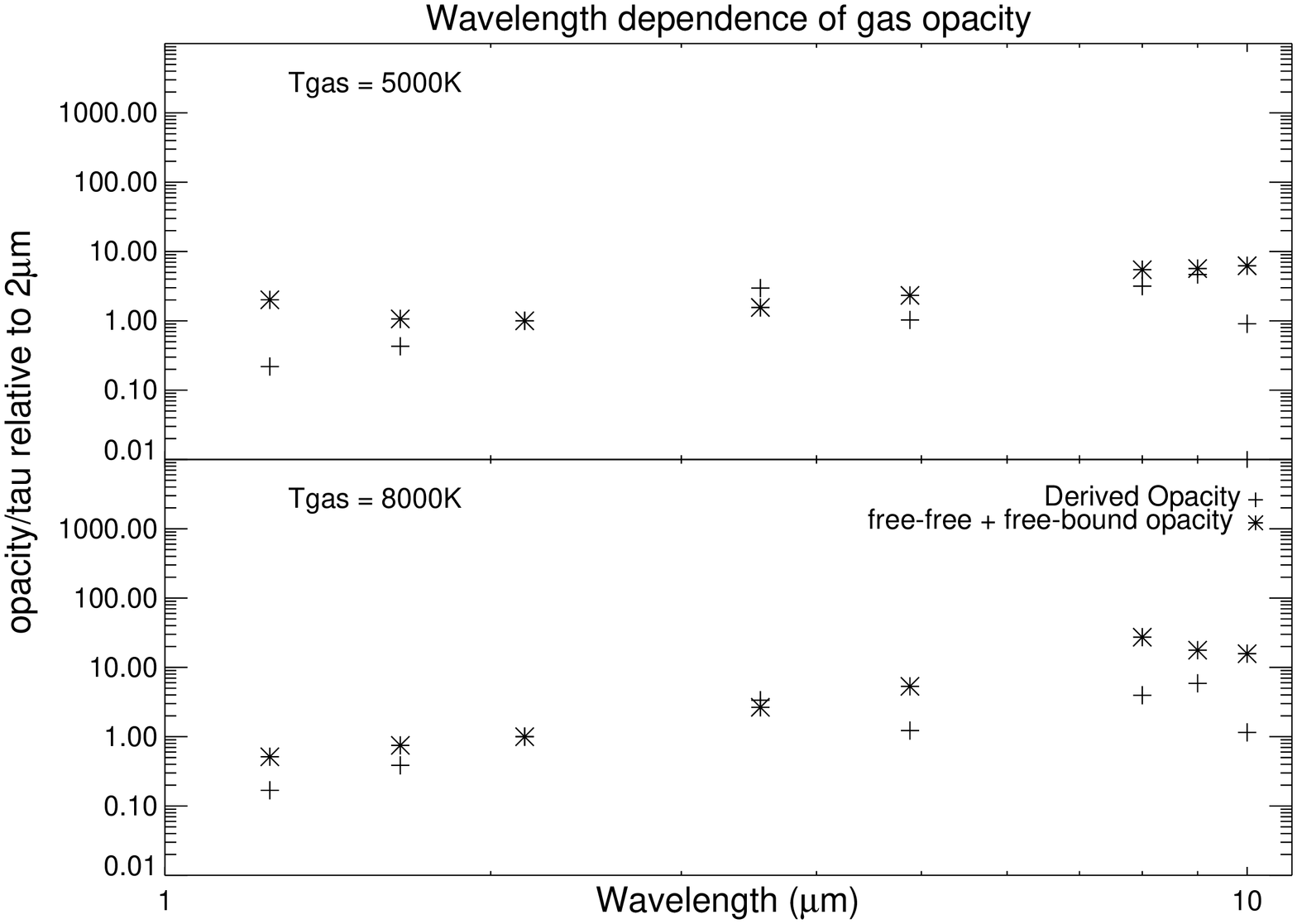}
}
\caption{The plusses (+)  represent  empirically derived gas opacities  from  observed photometry and NIR disk models for MWC275 (see Fig \ref{rim_atm} ). a) Top panel. The stars  represent fiducial theoretical molecular absorption opacities  smoothed over the photometry band for 2000K and 2500K gas respectively \citep{zhu2007}. The opacity jump at 5$\mu$m is due to water vapor.  b)  Bottom Panel. The gas absorption opacity at infrared wavelengths is dominated by 
free-free and free-bound transitions of H$^-$ at 5000K and by hydrogen at 8000K \citep{Ferguson2005, zhu2007}.
}
\label{opacity_curve}
\end{center}
\end{figure}

\begin{figure}[h]
\begin{center}
{
\includegraphics[angle=90,width=4.in]{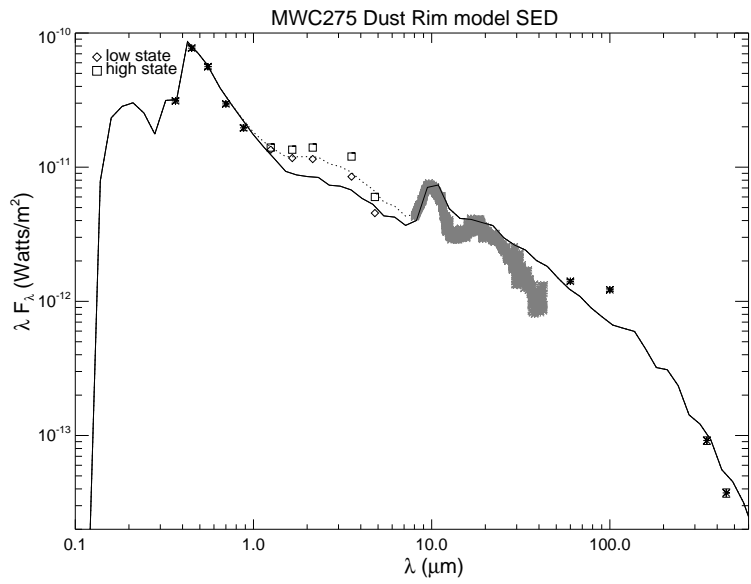}
}
\caption{MWC275 SED from UV to mm.The mid, far-infrared and sub-mm data are from \citet{Meeus2001} and references therein. The solid line traces the dust-disk model SED (see \S\ref{MIR_model_275}). The dotted line traces the dust-disk+smooth emission SED. The smooth component is modeled as optically-thin grey emission at 2500K.   The relative contributions of star, dust and gas to the total integrated flux are 0.79, 0.16 and 0.05 respectively. 
}
\label{MWC275_SED_total}
\end{center}
\end{figure}

\begin{figure}[h]
\begin{center}
{
\includegraphics[angle=90,width=4.0in]{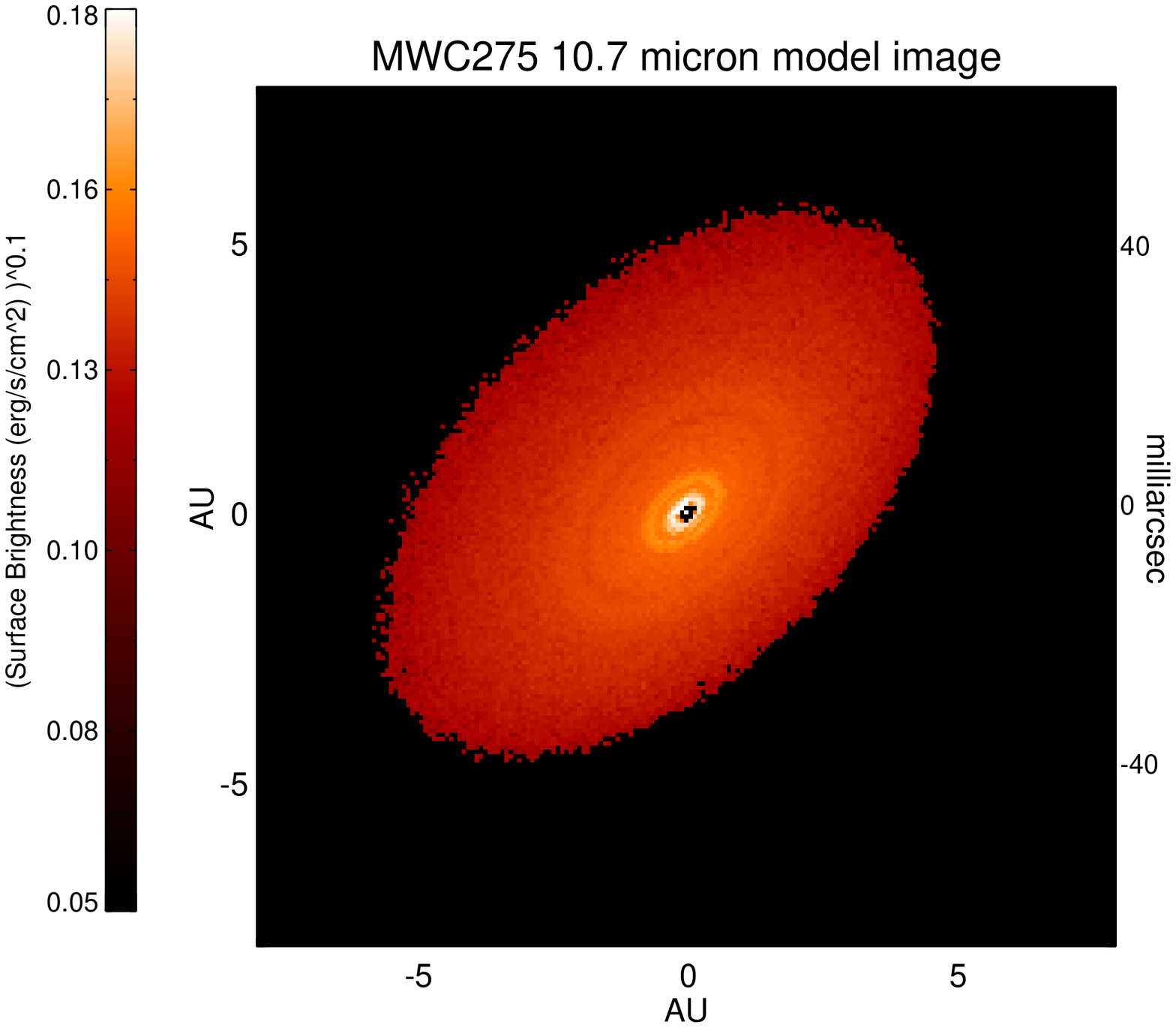}
\hphantom{.....}
\includegraphics[angle=90,width=3.0in]{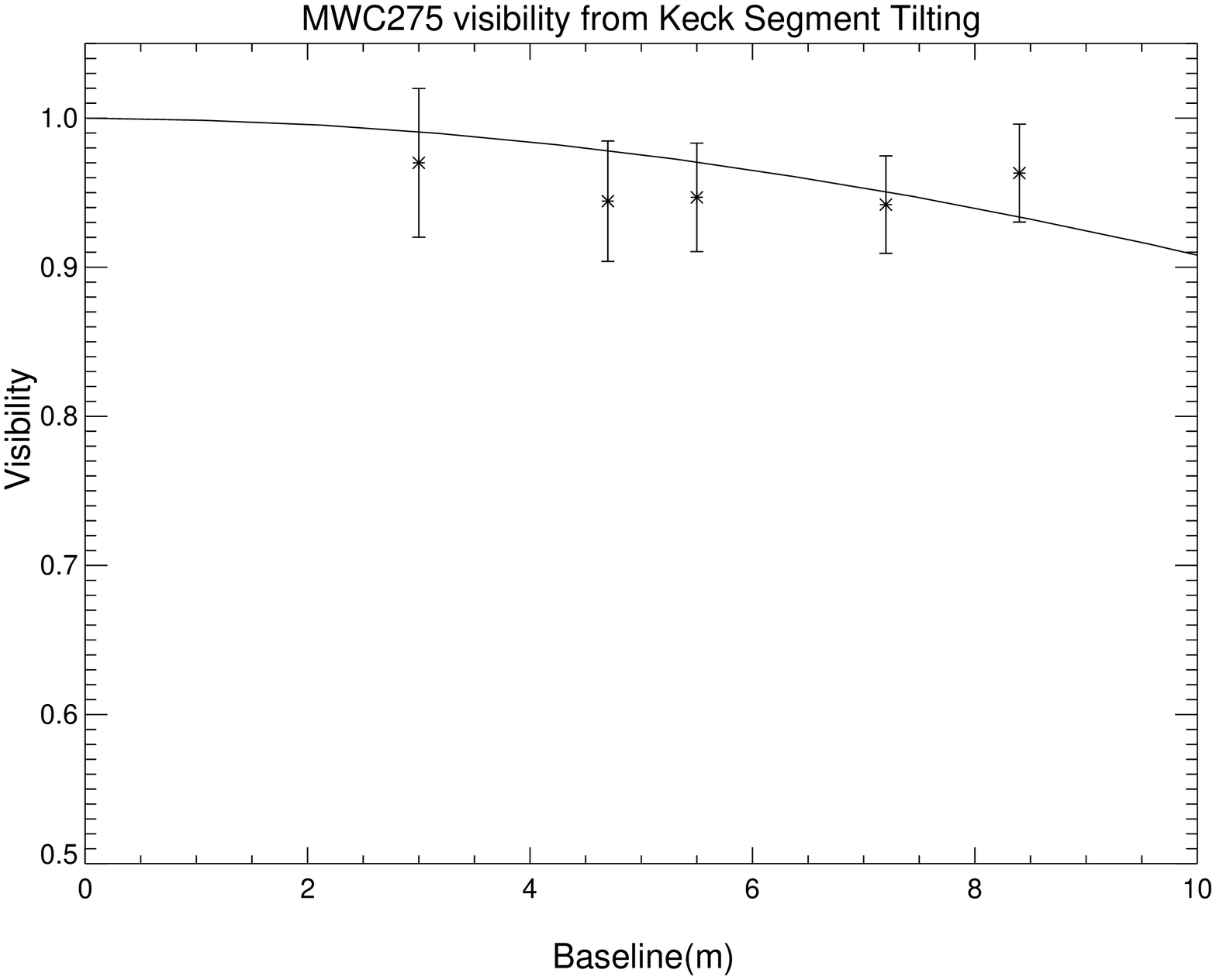}
\hphantom{.....}
\includegraphics[angle=90,width=3.0in]{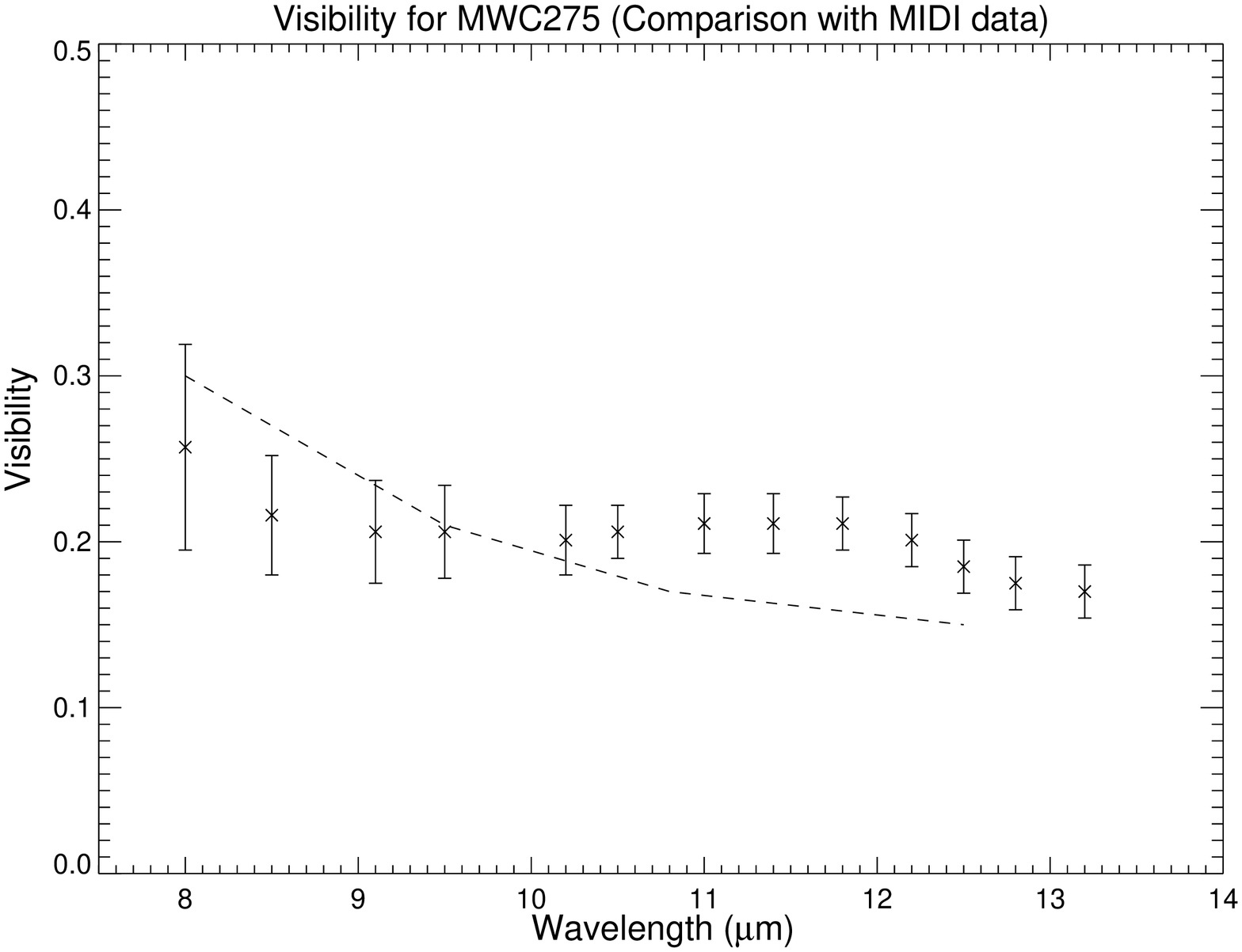}

}

\caption{MIR image and visibilities for MWC275. The disk has an inclination of  48$^o$ and a  PA of 136$^o$ (North is towards the  top and East  is on the left) a) Top panel. Synthetic 11$\mu$m TORUS image. b) Bottom left panel. Azimuthally  averaged 10.7$\mu$m visibilities from the Keck Segment Tilting Experiment \citep{monnier2008}. The `stars' are measured values and the solid line is the model visibility.  MWC275 is not resolved by Keck. c) Bottom right  panel. Model visibilities compared with MIDI \citep{Leinert2004} data. The MIDI data was obtained at a projected baseline of $\sim$ 99m and a PA of 16$^o$, nearly aligned with disk minor axis.
}
\label{MIDI}
\end{center}
\end{figure}

\begin{figure}[h]
\begin{center}
{
\includegraphics[angle=90,width=4.in]{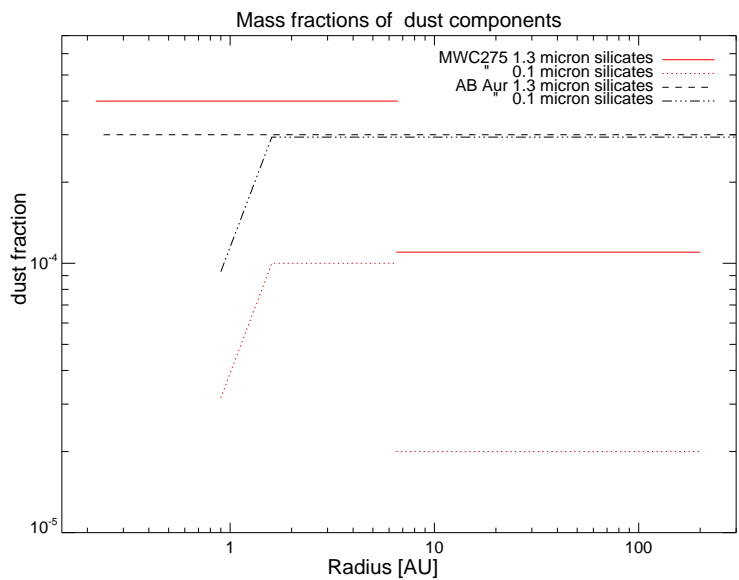}
}

\caption{Mass fractions of dust components relative to gas.  The micron and sub-micron grain fraction in MWC275 (red solid and dotted lines) have to be reduced  below 20\% of their values inside of 6.5AU at larger radii to fit the SED and interferometry. The silicate-grain opacities are from  \cite{ossen1992} and  the relative masses of  dust grains are  from \cite{boekel2005}. Between 0.9AU and 1.6 AU, 0.1$\mu$m grains are added smoothly  to avoid the formation of two distinct dust rims. \newline Bulk of the dust mass is in mm sized grains with a power law opacity profile \citep{Natta2004}. For AB~Aur, we also add a 50$\mu$m silicate component to improve SED fits between 40$\mu$m and 100$mu$m. The dust parameters are derived assuming gas and dust are well mixed.}

\label{dust_fraction}
\end{center}
\end{figure}

\begin{figure}[h]
\begin{center}
{
\includegraphics[angle=90,width=3.in]{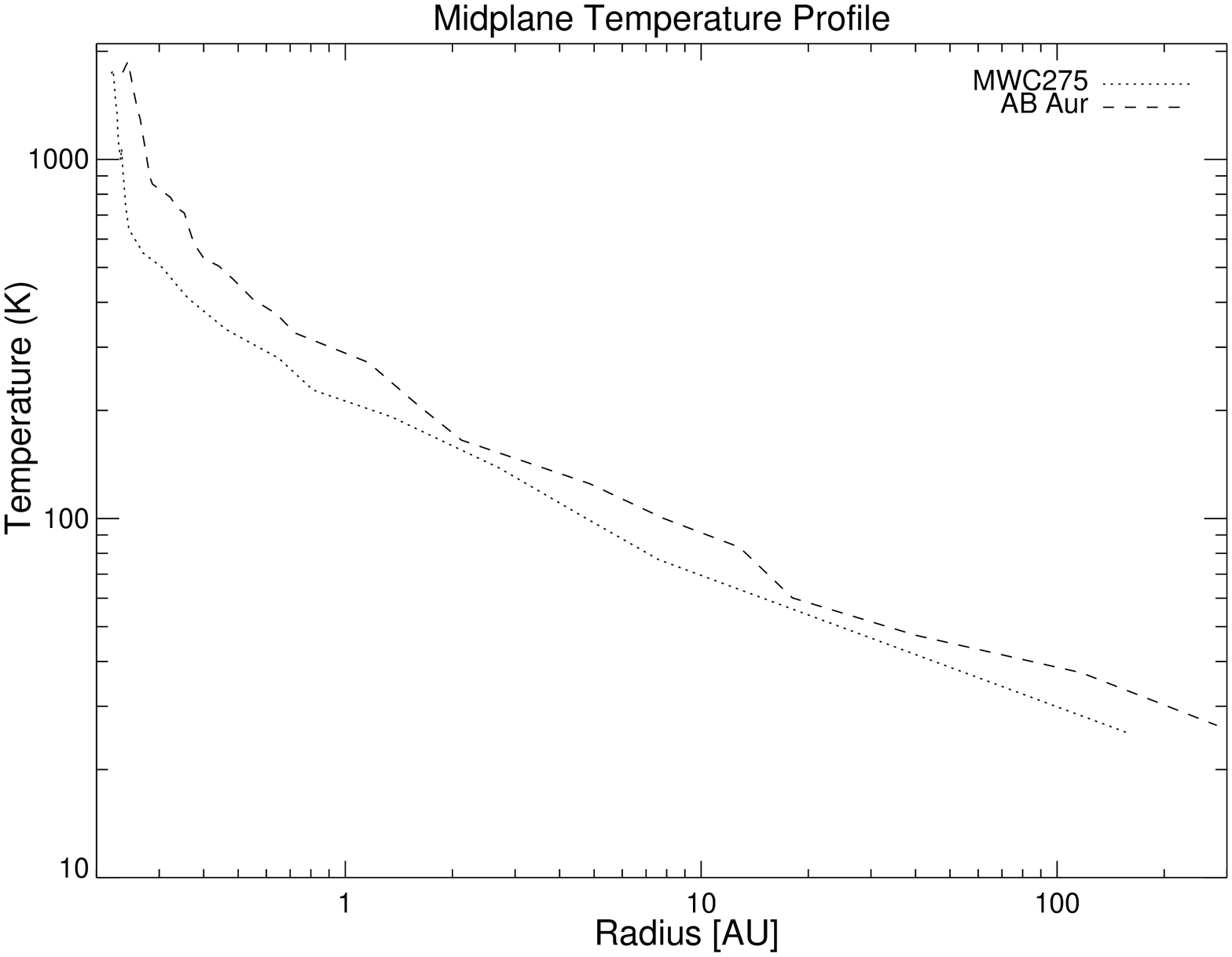}
\hphantom{.....}
\includegraphics[angle=90,width=3.in]{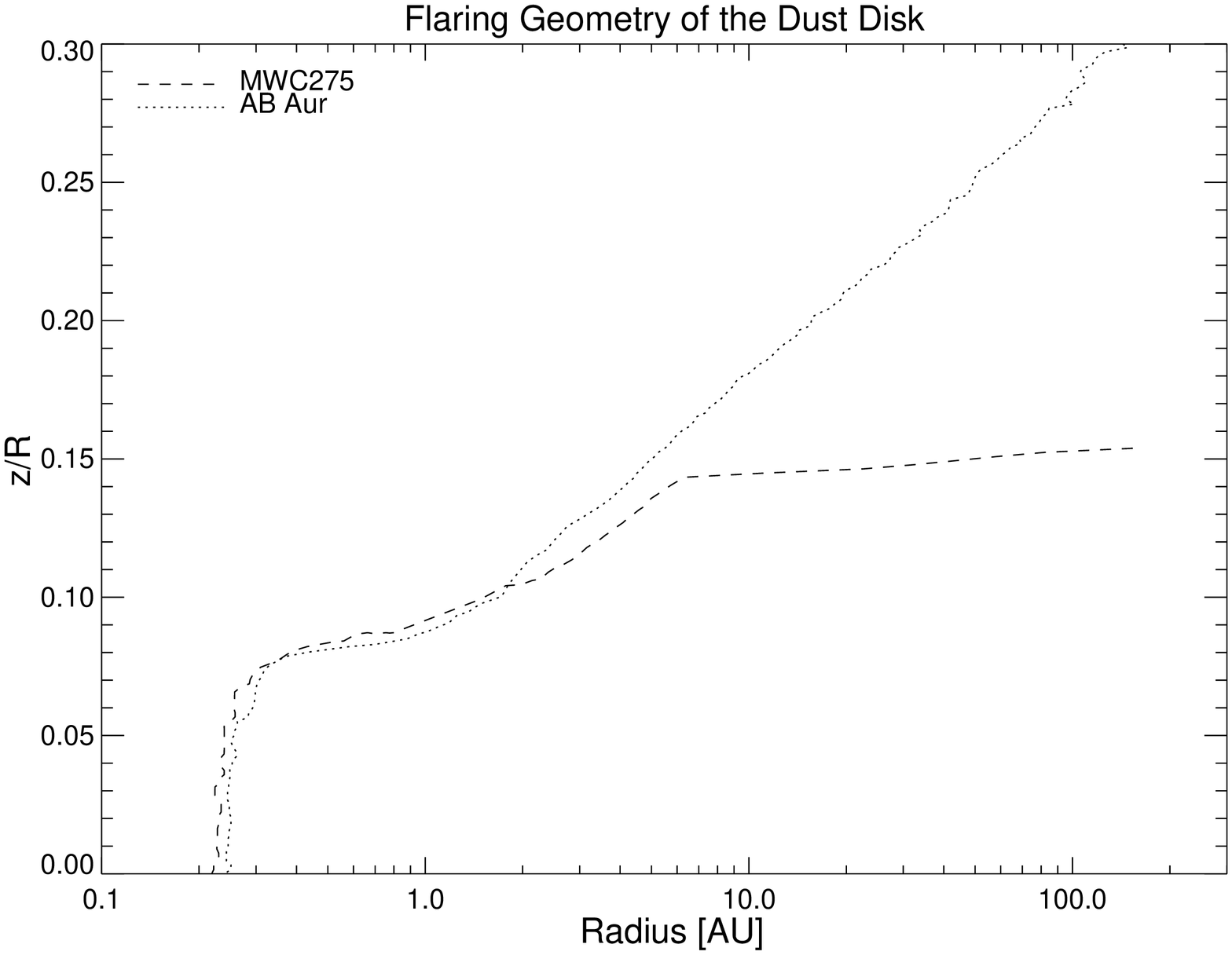}
}

\caption{Temperature profile and disk-surface shapes for MWC275 and AB~Aur.  a)  Left  panel. Midplane temperature profile for MWC275 (dotted line) and AB~Aur (dashed line). b) Right panel. The figure shows $\tau$=1 at 5500\AA~ surface of the disk  measured along radial lines from the central star. The y axis is the polar angle (0 is the equatorial plane)  in radians.
}
\label{profiles}
\end{center}
\end{figure}

\begin{figure}[h]
\begin{center}
{
\includegraphics[angle=90,width=3.in]{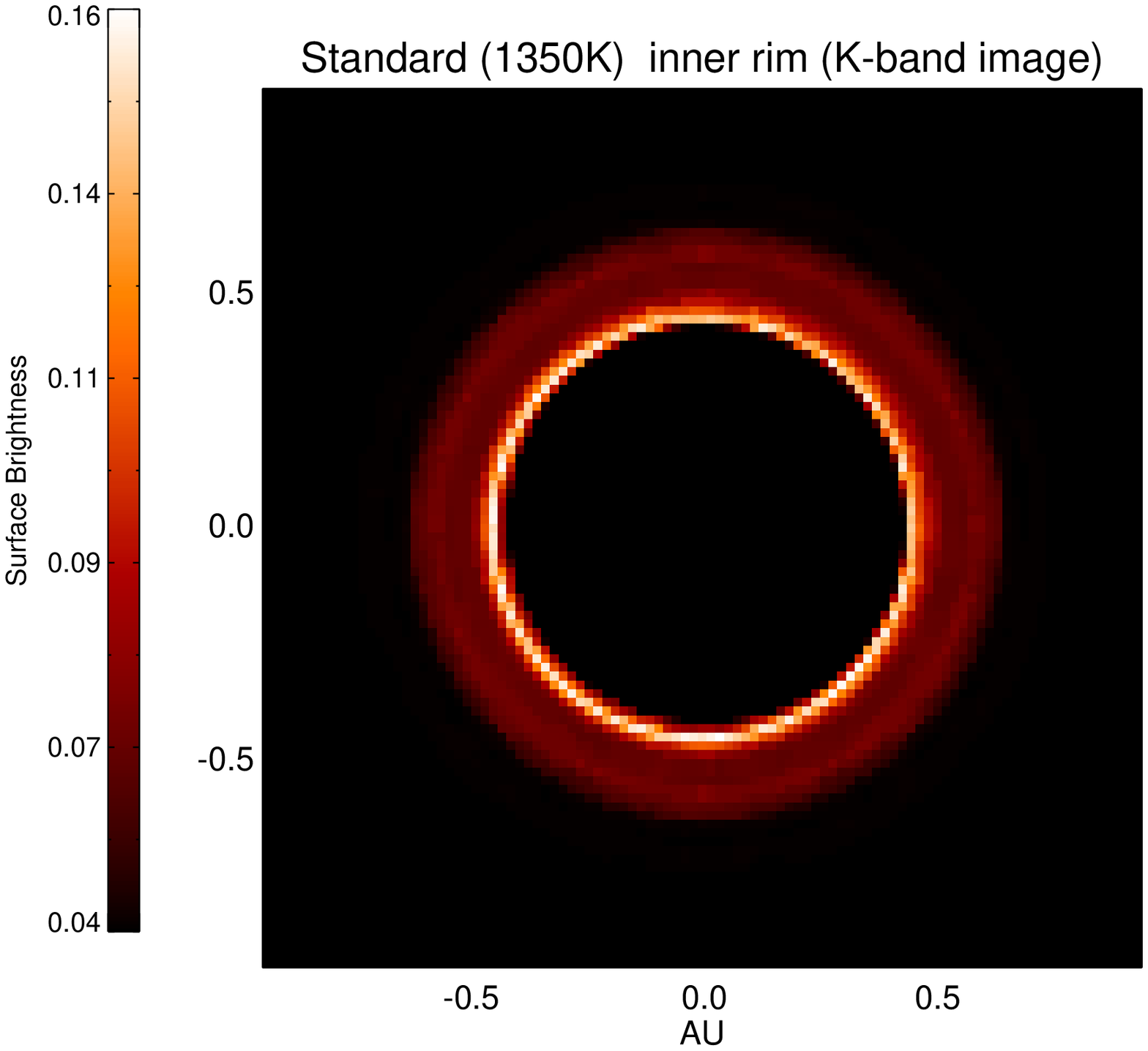}
\hphantom{.....}
\includegraphics[angle=90,width=3.in]{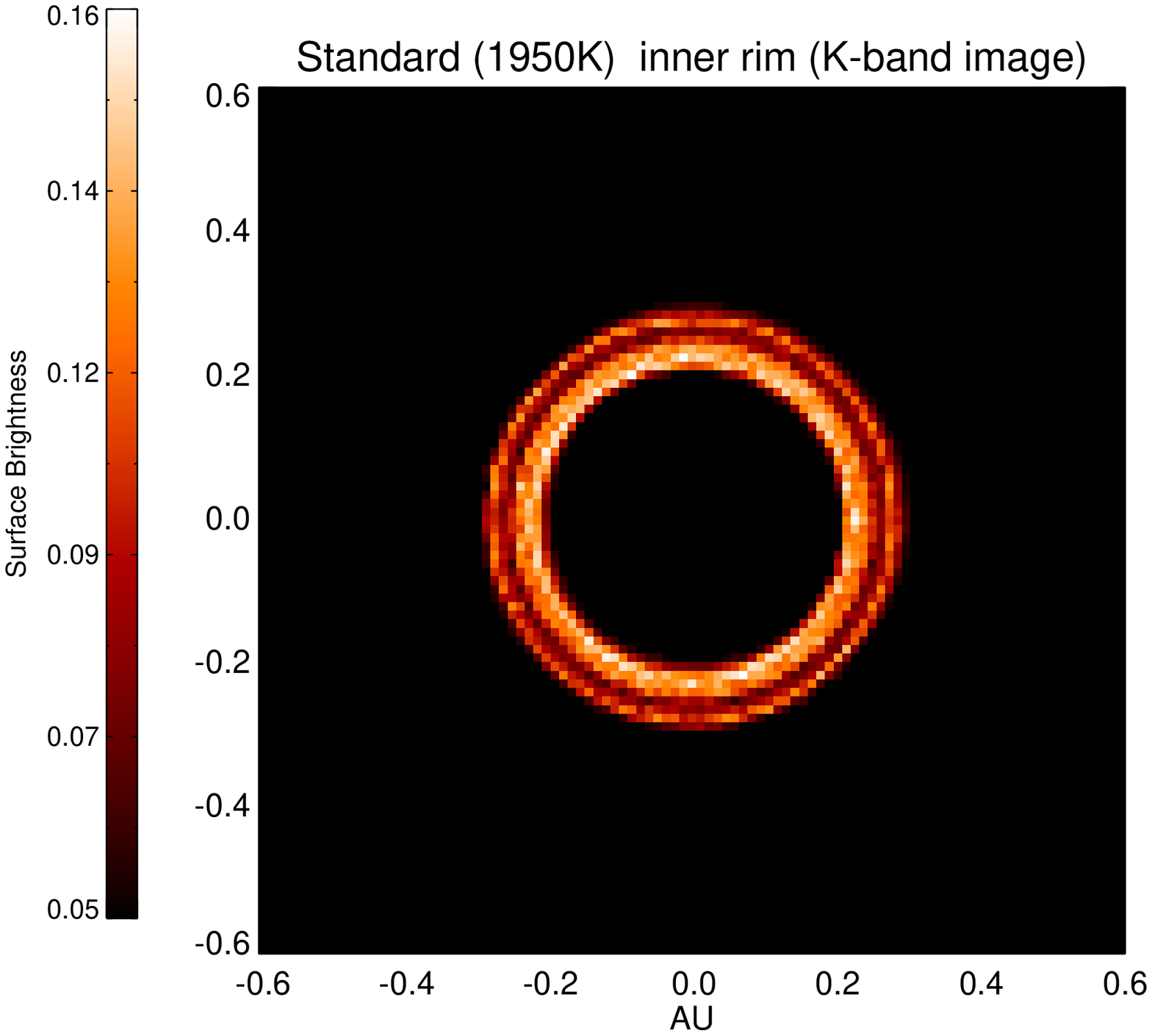}
\hphantom{.....}
\includegraphics[angle=90,width=3.in]{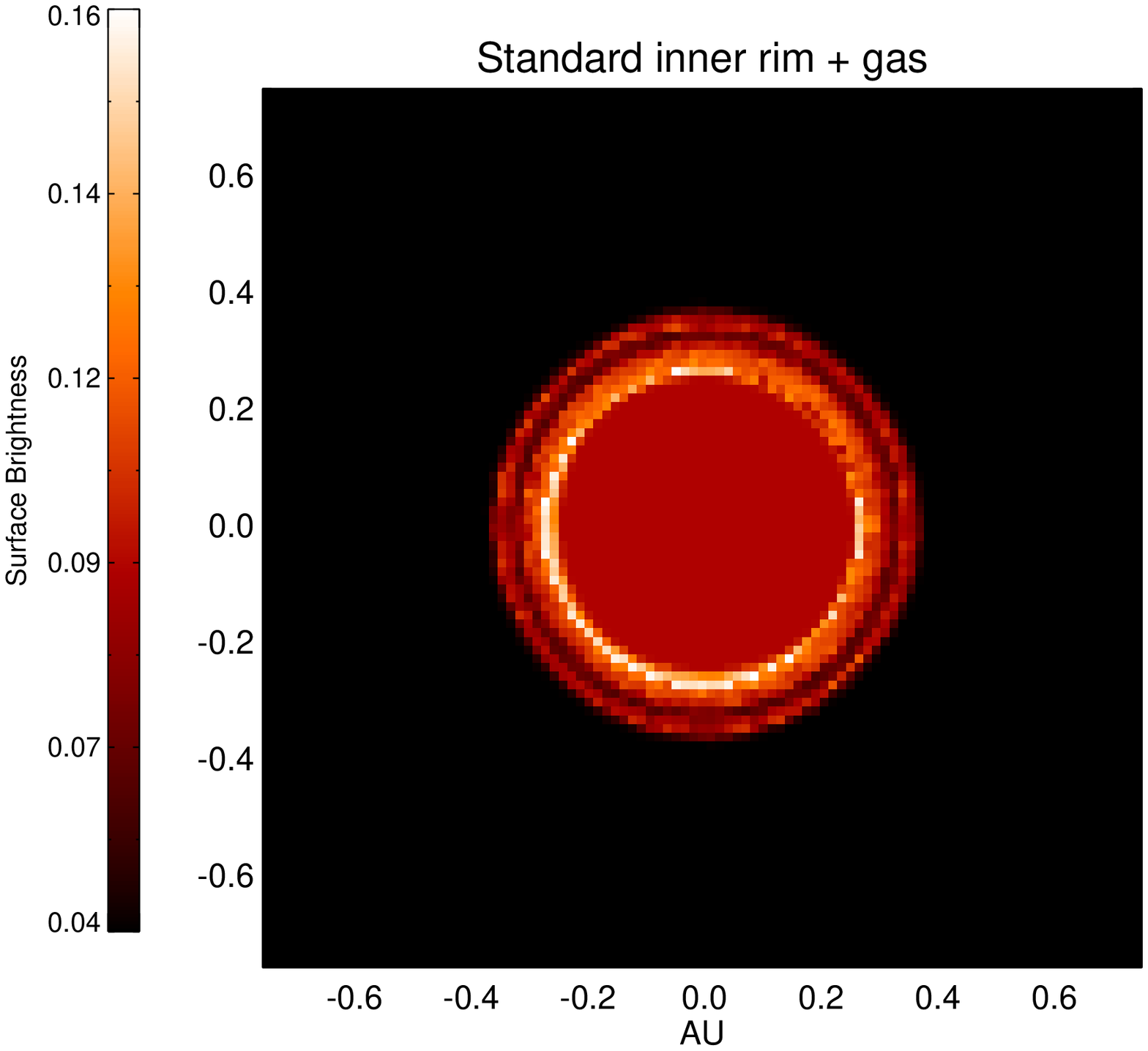}
}

\caption{Face-on  models for NIR emission in AB~Aur.  a) A  standard curved dust-rim-only model with rim-base temperature  $\sim$1350K.  b) Top  right  panel. Standard curved dust-rim-only model with rim-base temperature  $\sim$ 1950K.
c) Bottom panel. Curved dust-rim model with gas emission (modeled as a uniform disk centered on the star)  added inside  the dust rim in  to smooth out the emission profile. The central star has been suppressed in all the panels.
}
\label{rim_atm_faceon}
\end{center}
\end{figure}

\begin{figure}[t]
\begin{center}
{
\includegraphics[angle=90,width=5in]{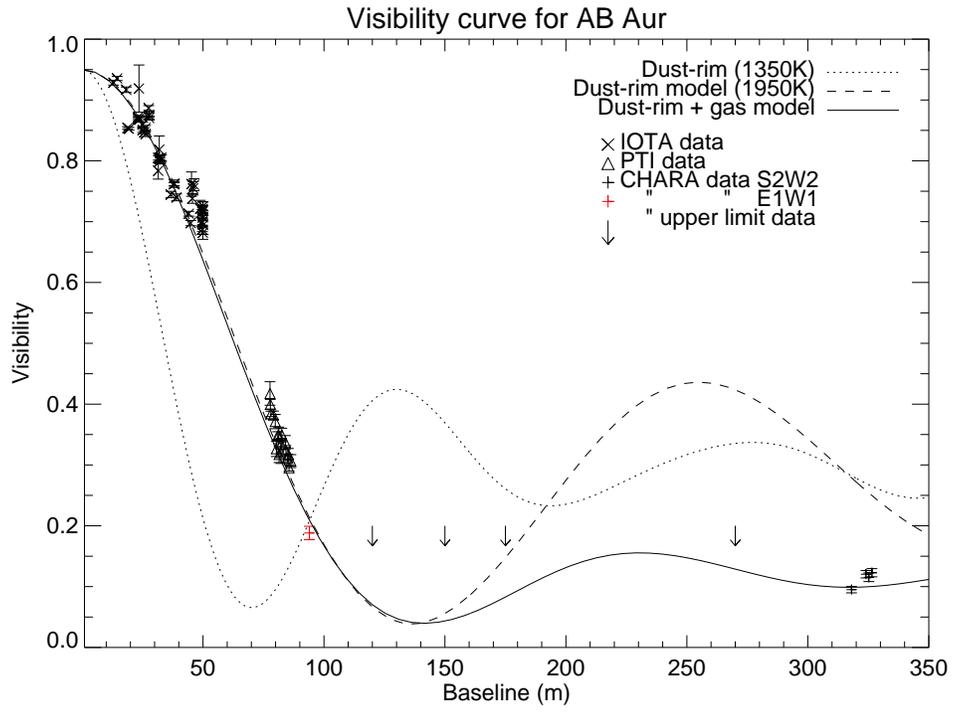}
}
\caption{AB~Aur visibility vs Baseline. The arrows are upper limits on the visibility. The quoted model temperatures are at the base of the dust rims. }
\label{ABAur_Vis_CHARA}
\end{center}
\end{figure}

\begin{figure}[h]
\begin{center}
{
\includegraphics[angle=90,width=5in]{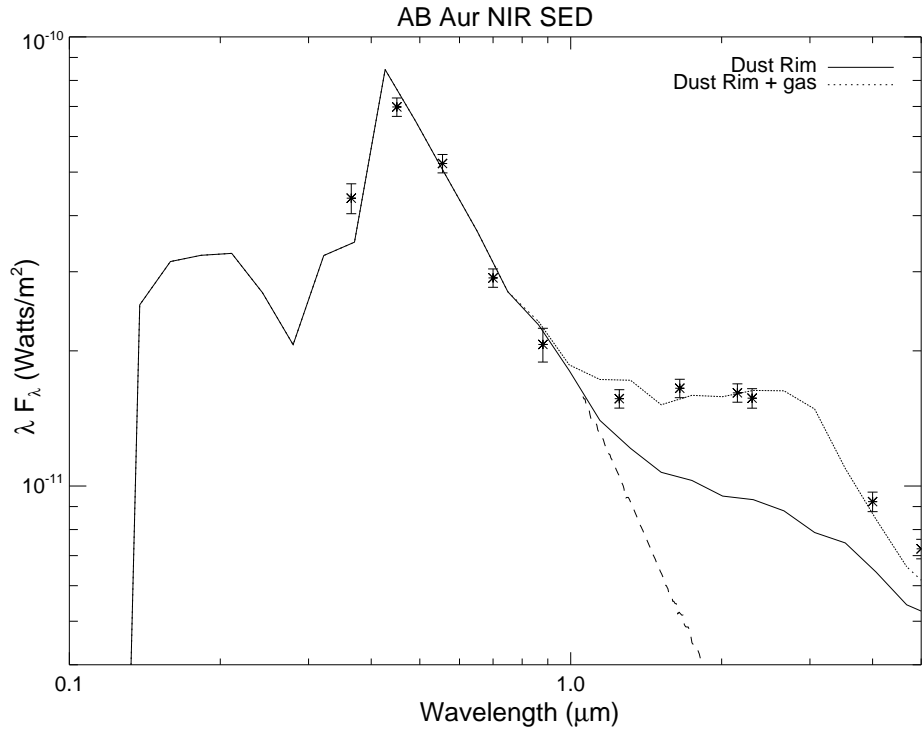}
}
\caption{The NIR SED for AB~Aur. The `stars' are photometry points from MDM (Table \ref{photometry_b}). The solid line is the SED produced by the `star + dust-rim only' model in Fig \ref{rim_atm_faceon}b. The dashed line traces the stellar SED. The dotted line includes emission from gas at 2500K, assuming that the gas opacity curve derived for MWC275 (see Fig. \ref{opacity_curve}) is valid for AB~Aur as well.
}
\label{ABAur_NIR_SED}
\end{center}
\end{figure}

\begin{figure}[h]
\begin{center}
{
\includegraphics[angle=90,width=3.0in]{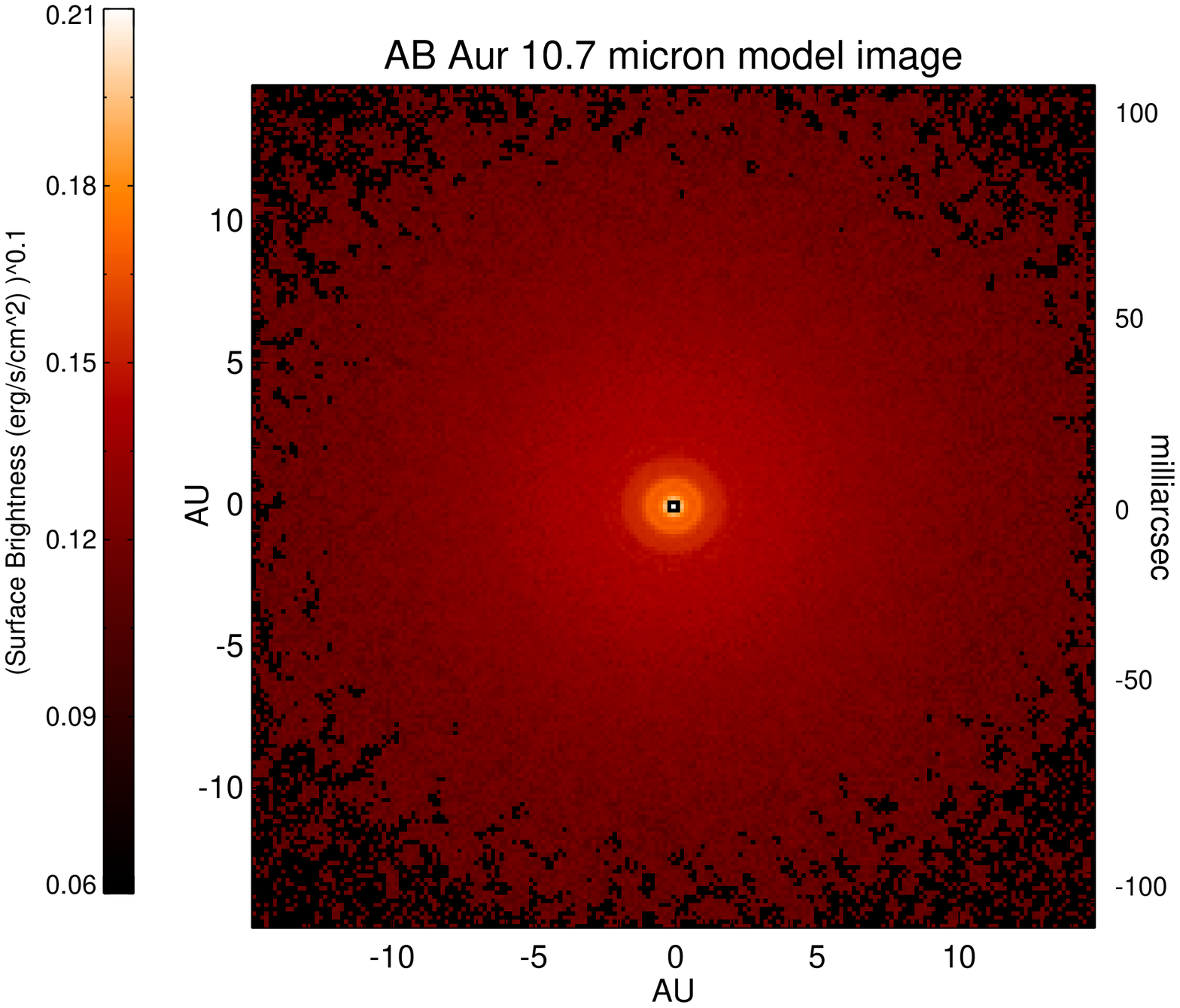}
\hphantom{.....}
\includegraphics[angle=90,width=3.0in]{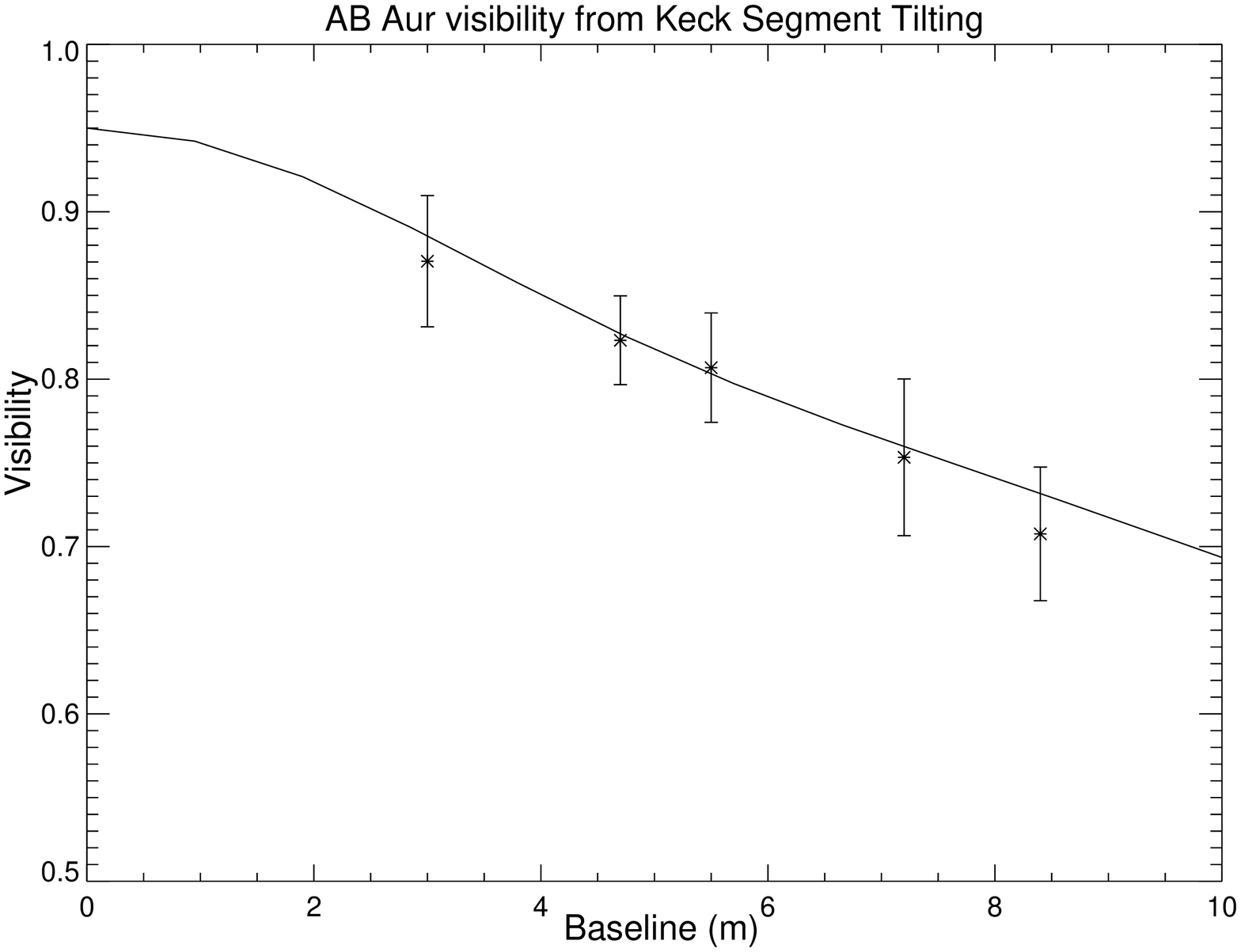}
}
\caption{10.7$\mu$m image and visibilities for AB~Aur. a) Left panel. Synthetic 10.7$\mu$m TORUS image  b)Right panel. Model visibilities (solid line) compared with azimuthally averaged Keck Segment Tilting data \citep{Monnier2004}. The model also includes 5\%  emission arising from an extended envelope.
}
\label{ABAur_LWS}
\end{center}
\end{figure}

\begin{figure}[h]
\begin{center}
{
\includegraphics[angle=90,width=4.in]{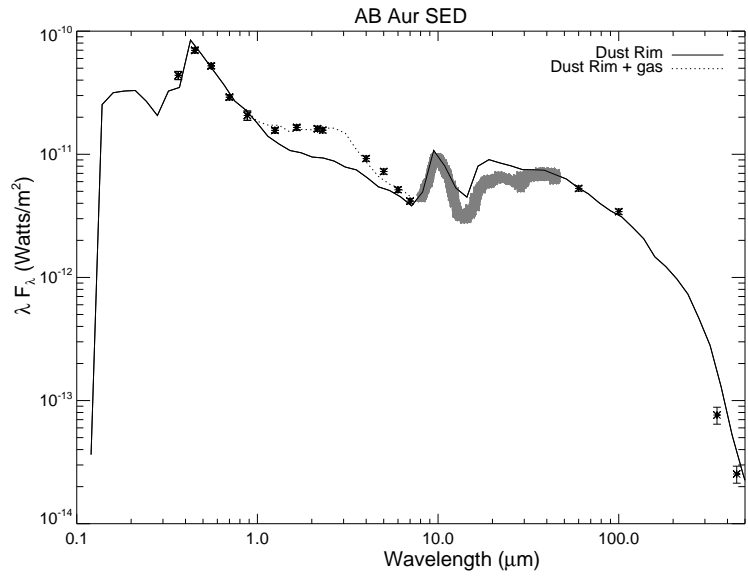}
}
\caption{AB~Aur SED from UV to mm. The mid, far-infrared and sub-millimeter data are from \citet{Meeus2001} and references therein. The solid line traces the dust-disk model SED (see \S\ref{MIR_model_ABAur}). The dotted line traces the dust-disk + gas model. The relative contributions of star, dust and gas to the total integrated flux are 0.67, 0.27 and 0.06 respectively. 
}
\label{ABAur_SED_total}
\end{center}
\end{figure}

\clearpage
\appendix
\section{Appendix}

\subsection{Photometry}
\begin{landscape}
\footnotesize
\begin{longtable}{ccccccccc}
\caption{UBVRI Photometry. Majority of the MDM targets are YSOs.} \label{UBVR}\\
\hline \\[-2ex]
\multicolumn{1}{c}{Target}  & \multicolumn{1}{c}{RA (J2000)} &  \multicolumn{1}{c}{Dec} &  \multicolumn{1}{c}{U} & \multicolumn{1}{c}{B} &\multicolumn{1}{c}{V} &
         \multicolumn{1}{c}{R} & \multicolumn{1}{c}{I}  & \multicolumn{1}{c}{UT Date} \\
        \multicolumn{1}{c}{}    &    \multicolumn{1}{c}{} &  \multicolumn{1}{c}{}  & \multicolumn{1}{c}{ } & \multicolumn{1}{c}{} &  \multicolumn{1}{c}{} & \multicolumn{1}{c}{}  &  \multicolumn{1}{c}{} &  \multicolumn{1}{c}{of Observation}\\ [-0.5ex] \hline \\ [-1.8ex] 
 \endhead
HIP 2080  & 00 26 16.5 & $+$ 03 49 33 &  -      & 6.72$\pm$0.05   &   6.81$\pm$0.04 &   6.94$\pm$0.06  & 6.93$\pm$0.04 & 08/27/2006  \\
HIP2370   &00 30 16.4  & $-$ 29 14 38 &  -      & 12.65$\pm$0.05 & 12.28$\pm$0.04 &  12.13$\pm$ 0.06 & 11.75$\pm$0.04 & ''\\
HIP3013   & 00 38 20.3 & $ -$14 59 54  &  -     & 10.67$\pm$0.05  &10.86$\pm$ 0.04&   6.95$\pm$ 0.06 &  11.07$\pm$0.04 &'' \\       
BP Tau      & 04 19 15.8 & $+$ 29 06 27 &13.20$\pm$ 0.15 &  13.36$\pm$0.03 & 12.32$\pm$0.04 & 11.45$\pm$0.04 & 10.60$\pm$0.04 & 11/29/2004 \\
CI Cam      & 04 19 42.1 & $+$55 59 57  &12.13$\pm$0.07 & 12.41 $\pm$0.06 & 11.77$\pm$0.05 & 10.79$\pm$ 0.05 & 9.99$\pm$0.04 & 11/27/2004\\       

                   &                         &                             &12.32$\pm$0.08  &12.70 $\pm$0.04 & 11.75$\pm$0.05 & 10.64$\pm$0.05  & 9.82$\pm$0.09 & 12/10/2005\\
                   &                         &                             & -                             & 12.59$\pm$0.05 & 11.71$\pm$ 0.03 & 10.72$\pm$ 0.04 & 9.94$\pm0.05$ & 08/28/2006\\ 
DG Tau     & 04 27 04.7   & $+$26 06 16 &13.93$\pm$0.04  &13.97$\pm$0.03  & 12.79$\pm$ 0.04 & 11.70$\pm$ 0.04 &   10.67$\pm$0.04 & 11/29/2004\\
DI Tau       &  04 29 42.5 & $+$26 32 49  &16.06$\pm$0.25  &14.45$\pm$ 0.03 & 12.96 $\pm$ 0.04 & 11.87$\pm$ 0.04 & 10.70$\pm$0.04 & ''\\
V830 Tau &  04 33 10.0  & $+$24 33 43  &14.66$\pm$0.15  &13.52$\pm$ 0.03 &12.21 $\pm$ 0.04  & 11.26$\pm$ 0.04 &  10.44$\pm$0.04 &''\\
LkCa 15    &  04 39 17.8 &$+$22 21 03  &13.98$\pm$0.07  &13.30$\pm$ 0.03 &12.09 $\pm$ 0.04 & 11.26$\pm$ 0.04 &  10.52$\pm$0.04 & " \\
GM Aur      &  04 55 11.0  &$+$30 21 59  & 13.90$\pm$0.04  &13.38$\pm$0.03 &12.19 $\pm$  0.04 &11.34$\pm$ 0.04 & 10.61$\pm$0.04 & '' \\
AB Aur      &  04 55 45.8  & $+$30 33 04  & 7.18 $\pm$0.08   &7.14 $\pm$0.04  & 7.01 $\pm$ 0.04  & 6.96 $\pm$  0.05 &  6.70$\pm$0.09 &12/10/2005 \\
                   &                          &                              &   -                          & 7.19 $\pm$0.05 & 7.05$\pm$ 0.05  & 7.01$\pm$ 0.06  &  6.80$\pm$0.04 & 08/27/2006 \\
MWC480   & 04 58 46.3   & $+$29 50 37   &  -                           & 7.91 $\pm$0.05 & 7.68$\pm$ 0.03 & 7.62$\pm$ 0.05  &   7.45$\pm$0.05 & 08/28/2006\\
UX Ori       &  05 04 30.0  &$ -$03 47 14    &10.93 $\pm$0.07 &10.70$\pm$0.06 &10.33$\pm$0.05 &10.07$\pm$ 0.05 &  9.75$\pm$0.04 & 11/29/2004\\
                   &                          &                               &10.43$\pm$ 0.08 &10.22$\pm$0.04 & 9.97$\pm$ 0.04 & 9.87$\pm$ 0.05 &  9.56$\pm$0.09 & 12/10/2005\\
RW Aur      & 05 07 49.5  & $ +$30 24 05  &10.86$\pm$0.04&11.07$\pm$ 0.03&10.32$\pm$ 0.04 & 9.78$\pm$ 0.04  &  9.17$\pm$0.04 &  11/29/2004\\
GW Ori      &  05 29 08.4   &$+$11 52 13    &11.60$\pm$ 0.04 & 11.29$\pm$0.03&10.16$\pm$ 0.04& 9.38$\pm$ 0.04 &  8.75$\pm$0.04 & 11/29/2004\\
MWC 758  &  05 30 27.5  &$ +$25 19 57    &8.59  $\pm$ 0.07 & 8.56 $\pm$0.06  &8.28$\pm$ 0.05 & 8.11$\pm$ 0.05  & 7.95$\pm$0.04 & 11/29/2004\\
                    &                          &                               &   -                          & 8.58 $\pm$ 0.05 &8.28$\pm$  0.03 & 8.15$\pm$ 0.04 & 7.92$\pm$0.05 & 08/28/2006\\
V380 Ori*   & 05 36 25.4   &$ -$06 42 58  & 10.29$\pm$0.07 & 10.43$\pm$0.06 & 10.06$\pm$0.05 & 9.67$\pm$ 0.05 &    8.99$\pm$0.04 &  11/29/2004\\
                    &                          &                             &10.07$\pm$ 0.08  &10.15$\pm$0.04  &9.74$\pm$ 0.04  &9.38 $\pm$ 0.05 &  8.80$\pm$0.09 & 08/27/2008\\
FU Ori         & 05 45 22.4   & $+$09 04 12  &   -                            &10.80$\pm$0.08  &9.52$\pm$ 0.15  & 8.63$\pm$ 0.05 &    8.11$\pm$0.15 &  11/29/2004\\
                     &                          &                             &11.57$\pm$0.08 &10.81$\pm$0.04  & 9.47$\pm$ 0.04 &  8.60$\pm$ 0.05 &  7.89$\pm$0.09 & 12/10/2005\\
                     &                          &                             &  -                            & 10.87$\pm$0.05 &9.65$\pm$ 0.03 & 8.87$\pm$  0.05 &   8.13$\pm$0.05 & 08/28/2006\\      
HD 45677   & 06 28 17.4   &$-$13 03 11   & 7.03$\pm$0.07    &   7.56$\pm$0.06 &7.51$\pm$ 0.05 & 7.33$\pm$  0.05 &   7.24$\pm$0.04 &11/27/2004\\     
                     &                          &                             & 6.99$\pm$0.08   &      -                         &7.52$\pm$  0.04& 7.35$\pm$  0.05 & 7.06$\pm$0.09 & 12/10/2005  \\ 
MWC 147   &  06 33 05.2   &  +10 19 20     & 8.45$\pm$0.07   &   8.87$\pm$0.06  &8.65$\pm$ 0.05 & 8.30$\pm$  0.05 &  8.02$\pm$0.04 &  11/27/2004\\
Z CMa         &  07 03 43.2   & -11 33 06       & 9.78$\pm$0.07   &   9.86$\pm$0.06  &9.17$\pm$ 0.05 & 8.50$\pm$  0.05  &   7.75$\pm$0.04 & ''\\
MWC 166   &  07 04 25.5    & -10 27 16       &7.08$\pm$0.07   &   7.55$\pm$0.06   &7.16$\pm$ 0.05 & 6.97$\pm$ 0.05 &   6.65$\pm$0.04 &''\\
HD 58647   & 07 25 56.1   &$-$14 10 44    & 6.66$\pm$0.07   &   6.874$\pm$0.06   &6.73$\pm$ 0.05 & 6.76$\pm$ 0.05 &  6.67$\pm$0.04 & ''\\
IRC+10216 & 09 47 59.4   &$+$13 16 44   &   -                           &    -                            &  $>$15.9                & 15.60$\pm$0.10&  12.52$\pm$0.04 &  ''\\        
Beta Leo     &11 49 03.6     &$+$14 34 19   &  -                           &   2.36$\pm$0.09    &2.18$\pm$0.08 & 2.17$\pm$0.09&  2.22$\pm$0.07 & 06/09/2006\\        
HD141569  &15 49 57.8    &$-$03 55 16     &  -                           &   7.32$\pm$0.09   &7.16$\pm$0.08 & 7.14 $\pm$0.09&  7.15$\pm$0.07 & "\\
                      &                        &                                &  -                            &  7.32$\pm$0.05   &7.14$\pm$0.04 & 7.23$\pm$ 0.06& 7.03$\pm$0.04 & 08/27/2006 \\ 
HD142666   &15 56 40.0  &$ -$22 01 40     &  -                            &  9.70$\pm$0.05    &9.05$\pm$0.04 & 8.77$\pm$0.06&   8.32$\pm$0.04 &  08/28/2006\\
HD143006   &15 58 36.9     &$-$22 57 15     & -                             &10.94$\pm$0.09   &10.12$\pm$0.08&9.63$\pm$0.09&  9.38$\pm$0.07 & 06/09/2006\\
                       &                        &                               &-                              & 10.95$\pm$ 0.05  &10.14$\pm$0.04&9.75 $\pm$0.06& 9.22$\pm$0.04 &08/28/2006\\
HD144432   &16 06 58.0  &$-$27 43 10     & -                             &  8.57 $\pm$0.09   & 8.21  $\pm$0.08& 7.98$\pm$0.09&  7.96$\pm$0.07 & 06/09/2006\\
                       &                        &                               & -                             &  8.62  $\pm$0.05   &8.20 $\pm$0.03 & 7.97$\pm$0.04 &  7.83$\pm$0.04 &08/28/2006\\
HD150193   &16 40 17.9   &$-$23 53 45     & -                             &  9.43 $\pm$0.09 & 8.87$\pm$0.08 & 8.48$\pm$0.09 &  8.20$\pm$0.07 & 06/09/2006\\
(MWC863)    &                        &                               &                               &                                &                             &                             &  \\ 
                        &                        &                              & -                              & 9.42 $\pm$0.05 &8.87$\pm$0.04   &8.58$\pm$0.06  & 8.04$\pm$0.04 & 08/27/2006\\
KKOph          & 17 10 08.0  &$-$27 15 18     & -                              & 12.77$\pm$0.09  &12.11$\pm$0.08&11.61$\pm$0.09 &  11.16$\pm0.07$ & 06/09/2006\\
HD158352   & 17 28 49.7  &$+$00 19 50    & -                          &5.67 $\pm$0.05 &  5.38   $\pm$0.04&  5.33$\pm$0.06  &   5.18$\pm$0.04 & 08/27/2006     \\
                             
HD158643   & 17 31 25.0   &$ -$23 57 45   &  -                              &4.81 $\pm$ 0.09 &  4.78  $\pm$0.08&  4.80$\pm$0.09 &  4.73$\pm$0.07 &  06/09/2006\\
                       &                          &   & -                       & 4.87 $\pm$0.05  &4.78    $\pm$0.03&  4.81 $\pm$0.04 &    4.62$\pm$0.04 &  08/28/2006\\
RSOph          & 17 50 13.2   &$-$06 42 28     &-                              &12.65 $\pm$ 0.08 &11.43 $\pm$0.06& 10.21$\pm$0.05 &  9.51$\pm$0.07 & 06/08/2006\\
MWC275       &   17 56 21.3 &$-$21 57 22   &  -                             &6.98  $\pm$0.08   & 6.84   $\pm$0.06&   6.86 $\pm$0.05 &  6.71$\pm$0.07 & ''\\
                        &                          &                             &  -                            &7.01  $\pm$0.05    & 6.86  $\pm$0.04&   6.90 $\pm$0.06 & 6.72$\pm$0.04 & 08/27/2006\\
HD169412    &    18 21 33.5 &$+$52 54 08  &   -                          &7.89   $\pm$0.09 &  7.82   $\pm$0.08 &    7.85$\pm$0.09  & 7.97$\pm$0.07 & 06/09/2006\\
MWC297       &   18 27 39.6    &$-$03 49 52   &  -                            &14.38 $\pm$0.09 &  12.26 $\pm$ 0.08& 10.29$\pm$0.09  & 9.10$\pm$0.07 & ''\\
                                                     &       &                      &   -                          &14.27 $\pm$ 0.05 &  12.23 $\pm$ 0.04& 10.30$\pm$0.06  & 8.92$\pm$0.04 & 08/27/2006\\
VVSer           & 18 28 47.9   &$+$00 08 40   &  -                            &13.23 $\pm$0.08 & 12.22   $\pm$0.06 & 11.36$\pm$0.05 &   10.61$\pm$0.07 &  06/08/2006\\
MWC 300     &  18 29 25.7   &$-$06 04 37    & -                             &12.87$\pm$0.09 & 11.82   $\pm$0.08&  11.01$\pm$0.09  & 10.56$\pm$0.07 & 06/09/2006\\
RCra             & 19 01 53.7    &$-$36 57 08    & -                              &12.17$\pm$0.05 & 11.45   $\pm$0.03&  10.82$\pm$0.05  & 10.15$\pm$0.05 & 08/28/2006\\
TCra              &  19 01 58.8   &$-$36 57 50    & -                              &13.67$\pm$0.05 & 12.50   $\pm$0.03&  11.87$\pm$0.05  &  11.23$\pm$0.05 & ''\\
MWC614      &    19 11 11.3  &$+$15 47 16   &  -                           & 7.50 $\pm$0.05  & 7.37   $\pm$0.04 &    7.41$\pm$0.06    &   7.26$\pm$0.04 & 08/27/2006\\
HIP96720     &   19 39 41.4  & $+$14 02 53   &  -                           &11.06$\pm$0.05 & 10.65  $\pm$ 0.03&  10.38$\pm$0.05   &  10.06$\pm$0.05 & 08/28/2006\\
V1295 Aql     &  20 03 02.5   &$+$05 44 17  &   -                           & 7.85 $\pm$0.08  & 7.79   $\pm$0.06  &   7.69$\pm$0.05     &   7.59$\pm$0.07 & 06/08/2006\\
                        &                            &                            &-                               & 7.93$\pm$0.05  & 7.80   $\pm$0.04 &    7.77$\pm$0.06  &  7.58$\pm$0.04 &  08/27/2006\\
V1685Cyg     &   20 20 28.2 &$+$41 21 52 & -                              &11.63$\pm$0.09 & 10.79  $\pm$ 0.08&  10.06$\pm$0.09  &   9.58$\pm$0.07 &   06/09/2006\\
MWC342    &   20 23 03.6  & $+$39 29 50   &  -                             & 11.88$\pm$0.09      & 10.57$\pm$0.08 &   9.44$\pm$0.09   &   8.72$\pm$0.07 &  ''\\
V1057 Cyg    &   20 58 53.7  &$+$44 15 28  & 16.00$\pm$0.28 & 14.32$\pm$0.06 & 12.50  $\pm$ 0.05&  11.18$\pm$ 0.05  &   9.78$\pm$0.04 & 11/27/2004\\
MWC361       &  21 01 36.9  &$+$68 09 48    & -                             & 7.75$\pm$0.05  & 7.33     $\pm$0.03 &   7.00$\pm$0.04     &  6.62$\pm$0.05 & 08/28/2006\\
AS 477          &  21 52 33.9   &$+$47 13 38  & 10.48$\pm$ 0.07  &10.52$\pm$ 0.06 & 10.07 $\pm$0.05 &  9.85$\pm$0.05   &  9.50$\pm$0.04 &  11/27/2004\\
HIP113937   &    23 04 23.6 &$-$24 06 56   & -                               & 9.12$\pm$0.05 &  9.07   $\pm$0.04  &  9.18$\pm$0.06    &     9.08$\pm$0.04 & 08/27/2006\\
HIP114547   &    23 12 09.4 &$-$25 24 14   &  -                              & 9.51$\pm$0.05  & 9.12   $\pm$0.04 &   8.98$\pm$ 0.06    &    8.67$\pm$0.04 & ''\\
HIP115858   &   23 28 25.2  &$-$25 25 14   & -                              & 7.09$\pm$0.05  & 6.88   $\pm$0.04  &  6.92$\pm$ 0.06     &     6.72$\pm$0.04 &  ''\\
\hline
\end{longtable}
\end{landscape}
\clearpage

\begin{landscape}
\footnotesize
\begin{longtable}{ccccccc}
\caption{JHK Photometry. Majority of the MDM targets are YSOs.} \label{IJHK}\\
\hline \\[-2ex]
\multicolumn{1}{c}{Target}  & \multicolumn{1}{c}{RA (J2000)} &  \multicolumn{1}{c}{Dec} &  \multicolumn{1}{c}{J} & \multicolumn{1}{c}{H} &\multicolumn{1}{c}{K} &
          \multicolumn{1}{c}{UT Date} \\
\multicolumn{1}{c}{ } &    \multicolumn{1}{c}{ } & \multicolumn{1}{c}{} &  \multicolumn{1}{c}{} & \multicolumn{1}{c}{}  &  \multicolumn{1}{c}{} &  \multicolumn{1}{c}{of Observation}\\ [-0.5ex] \hline \\ [-1.8ex] 
 \endhead
V892Tau  &  04 18 40.6&$+$28 19 16&       8.61$\pm$0.05 &  7.08$\pm$0.05 &   5.86$\pm$0.05 & 12/17/2005   \\
BP Tau     &  04 19 15.8 &$+$29 06 27 &     9.10$\pm$0.10 &  8.37$\pm$0.10 &  7.90$\pm$0.10 & 12/01/2004    \\ 
                   &                      &                       &  9.05$\pm$0.05 &  8.32$\pm$0.05&   7.85$\pm$0.05  & 12/17/2005   \\
                                     
CI Cam     &  04 19 42.1& $+$55 59 58 &  7.20$\pm$0.10 &  5.68$\pm$0.10 &  4.44$\pm$0.10   &  12/01/2004 \\
                   &                        &                       & 7.01$\pm$0.05  & 5.63$\pm$0.05 &  4.35$\pm$0.05 &12/17/2005      \\
                                                                
DG Tau     &  04 27 04.7&$+$26 06 16  &  8.92$\pm$0.05  & 7.95$\pm$0.05 &  7.15$\pm$0.05   & 12/17/2005   \\

V830 Tau &  04 33 10.0& $+$24 33 43 &   9.36$\pm$0.05 &  8.73$\pm$0.05 &  8.52$\pm$0.05   & 12/17/2005    \\
                      
LkCa 15    & 04 39 17.8&$+$22 21 04  & 9.33$\pm$0.05 &  8.72$\pm$0.05 &  8.23$\pm$0.05 & 12/17/2005    \\ 
                                    
GM Aur     & 04 55 11.0&$+$30 22 00  &   9.40$\pm$0.05 &  8.80$\pm$0.05 &  8.52$\pm$0.05 & 12/17/2005      \\
AB Aur      & 04 55 45.8&$+$30 33 04   & 5.99$\pm$0.05  & 5.28$\pm$0.05  & 4.37$\pm$0.05    & 12/17/2005  \\
                    
MWC480  & 04 58 46.3&$+$29 50 37.0  &  6.90$\pm$0.05 &  6.38$\pm$0.05  & 5.57$\pm$0.05  & 12/17/2005     \\
 RW Aur     &  05 07 49.5&$+$30 24 05  &    8.22$\pm$0.05 &  7.67$\pm$0.05  & 7.16$\pm$0.05  & 12/01/2004   \\  
                    &                      &                         &   8.34$\pm$0.10 & 7.66$\pm$0.10  & 7.18$\pm$0.10   & 12/17/2005  \\
                   
GW Ori     &  05 29 08.4&$+$11 52 13   &    7.42$\pm$0.05  & 6.67$\pm$0.05 & 5.83$\pm$0.05 & 12/17/2005       \\
MWC 758  & 05 30 27.5&$+$25 19 57  &    7.20$\pm$0.10  & 6.55$\pm$0.10 & 5.80$\pm$0.10 & 12/01/2004 \\
                     &                           &                  &    7.26$\pm$ 0.05 &  6.70$\pm$0.05&  5.92$\pm$0.05  & 12/17/2005     \\
MWC120   &  05 41 02.3& $-$02 43 00 &     7.24$\pm$0.05 &  6.66$\pm$0.05 & 5.77$\pm$0.05  & 12/17/2005     \\
FU Ori       &   05 45 22.4&  $+$09 04 12&    6.55$\pm$0.05 &  5.89$\pm$0.05 & 5.32$\pm$0.05  & 12/17/2005   \\
HD 45677 &  06 28 17.4& $-$13 03 11 &  6.85$\pm$0.05  &  6.22$\pm$0.05 & 4.61$\pm$0.05  & 12/17/2005       \\   
MWC 147  &  06 33 05.2& $+$10 19 20&  7.34$\pm$0.05 &  6.70$\pm$0.05 &  5.73$\pm$0.05 & 12/17/2005     \\  
                    
Z CMa        &  07 03 43.2& $-$11 33 06 &  6.66$\pm$0.05  & 5.45$\pm$0.05 & 3.94$\pm$0.05 & 12/17/2005       \\    
                  
MWC 166  &  07 04 25.5& $-$10 27 16 &  6.32$\pm$0.05 &  6.32$\pm$0.05 &  6.27$\pm$0.05 & 12/17/2005    \\

Beta Leo      &  11 49 03.6& $+$14 34 19 &   1.92$\pm$0.08 &  1.96$\pm$0.07 & 1.90$\pm$0.08 & 06/02/2006  \\
                       
LambdaVir   &  14 19 06.6& $-$13 22 16 &    4.26$\pm$0.08 &  4.25$\pm$0.07 & 4.20$\pm$0.08  & ''     \\
HD141569   &  15 49 57.8&  $-$03 55 16&  6.67$\pm$0.08  & 6.54$\pm$0.07 & 6.48$\pm$0.08 & 06/02/2006\\ 
                       
HD143006  &  15 58 36.9   & $-$22 57 15 &      8.18$\pm$0.08 &  7.57$\pm$0.07 & 5.94$\pm$0.08 &  06/02/2006    \\
HD144432  &  16 06 58.0 & $-$27 43 10&          7.23$\pm$0.08   &  6.69$\pm$0.07 & 6.14$\pm$0.08 & 06/02/2006      \\
                      
HD150193  &  16 40 17.9  &$-$23 53 45&    6.84$\pm$0.08 &  6.02$\pm$0.07 & 5.13$\pm$0.08   &   06/02/2006  \\
(MWC863)   &                        &                      &                               &                              &                               &    \\
                                                       
KKOph         &  17 10 08.1 & $-$27 15 18&  8.46$\pm$0.08 &  7.09$\pm$0.07 & 5.71$\pm$0.08  & 06/02/2006     \\
HD158643  &  17 31 25.0 & $-$23 57 46 &   4.76$\pm$0.08 &  4.63$\pm$0.07 & 4.34$\pm$0.08 & 06/02/2006     \\  
          
RSOph         &  17 50 13.2 &$-$06 42 29  &   7.94$\pm$0.08  & 7.17$\pm$0.07 & 6.77$\pm$0.08  & 06/04/2006     \\ 
                        
MWC275      &  17 56 21.3 &$-$21 57 22 &6.20$\pm$0.08  & 5.48$\pm$0.07 & 4.59$\pm$0.08  & 06/02/2006     \\     
                  
HD169412   &  18 21 33.5  &$+$52 54 08& 7.77$\pm$0.08  & 7.78$\pm$0.07 & 7.79$\pm$0.08 & 06/02/2006      \\
MWC297      &  18 27 39.6    &$-$03 49 52 &  6.06$\pm$0.08  & 4.54$\pm$0.07 & 3.12$\pm$0.08 & 06/04/2006      \\

VVSer          & 18 28 47.9   &$+$00 08 40& 8.60$\pm$0.08  & 7.37$\pm$0.07 & 6.20$\pm$0.08 & 06/02/2006 \\   
                      
MWC614    &  19 11 11.3   &$+$15 47 16 &    6.91$\pm$0.08  & 6.58$\pm$ 0.07 & 5.88$\pm$0.08 & 06/03/2006 \\                     
V1295Aql  &    20 03 02.5  &$+$05 44 17&   7.15$\pm$0.08  & 6.61$\pm$0.07 & 5.75$\pm$0.08     & 06/02/2006   \\
                 
V1685Cyg &    20 20 28.3 & $+$41 21 52& 7.97$\pm$0.05 &  7.01$\pm$0.05&  5.86$\pm$0.05 & 12/17/2005      \\                
                   &                          &                        &  7.93$\pm$0.10 &  6.96$\pm$0.04 & 5.81$\pm$0.07 & 06/03/2006      \\
MWC342    &   20 23 03.6  & $+$39 29 50   &   7.01$\pm$0.05  & 5.98$\pm$0.05 & 4.78$\pm$0.05  & 12/17/2005   \\           
                      &                        &                            &  6.94$\pm$0.08 &  5.92$\pm$0.08 & 4.65$\pm$0.08 & 06/03/2006  \\
MWC361    &    21 01 36.9  &$+$68 09 48 &   6.12$\pm$0.05 &  5.58$\pm$0.05 & 4.77$\pm$0.05 & 12/17/2005       \\ 
MWC1080   &  23 17 25.6   &$+$60 50 43&    7.38$\pm$0.05 &  6.04$\pm$0.05 & 4.68$\pm$0.05   & 12/17/2005     \\

\hline
\end{longtable}
\end{landscape}
\clearpage

\subsection{``Effective Baselines'' as a tool in characterizing visibility information on MWC275.}
\label{appendix2}
Let $B_{projected}$ be the projected interferometric baseline and  let V($B_{projected}$) be the visibility for a circularly symmetric brightness distribution. For a flat disk inclined at angle $\phi$ and oriented at some PA, we plotted V($B_{eff}$) in Fig \ref{MWC275_Vis_CHARA}. The effective baseline $B_{eff}$ is defined as -
 $$B_{eff} = B_{projected} \sqrt{{\rm cos}^2(\theta) + {\rm cos}^2(\phi){\rm sin}^2(\theta)}  $$
where $\theta$ is the angle between the uv vector for the observation and the major axis of the inclined disk and $\phi$ is the inclination of the disk (0$^{o}$ inclination is face on). Effective baselines account for the decrease in interferometric resolution due to the inclination of the disk in the sky. They capture the geometry of flat disks correctly, but the geometry of finitely thick disks is represented  only approximately (optical depth effects and 3-D geometry of thick disks are not taken into account). Here, we argue that effective baselines are good (albeit approximate) tools for capturing details of the MWC275 disk geometry.

\begin{figure}[h]
\begin{center}
{
\includegraphics[angle=90,width=3.7in]{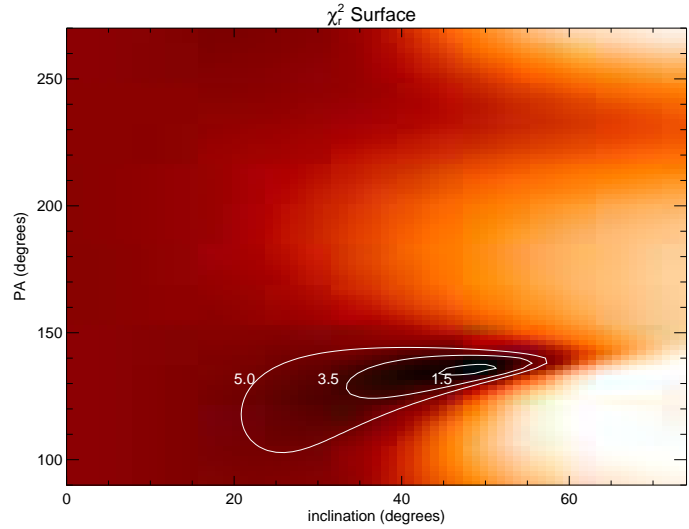}
}
\caption{Reduced $\chi^{2}$ surface for the linear fits to the observed visibilities (obtained with the W1W2, S2W2 and E2S2 CHARA-telescope-pairs) as a function of effective interferometric baseline. The solid curves are reduced $\chi^2$ contours of 5, 3.5 and 1.5 respectively.
}
\label{chi2}
\end{center}
\end{figure}

\begin{figure}[h]
\begin{center}
{
\includegraphics[angle=90,width=3.7in]{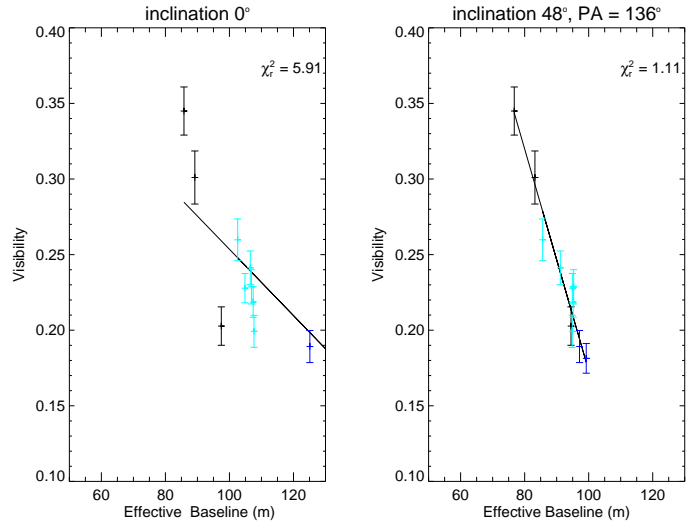}

}

\caption{ Linear fits to the observed visibilities (obtained with the W1W2, S2W2 and E2S2 CHARA-telescope-pairs) as a function of effective interferometric baseline. The data symbols are explained in Fig. 3a.
}
\label{PA_inc}
\end{center}
\end{figure}

In order to determine the inclination angle and sky orientation of the disk, we adopted the following procedure. MC275 visibility values  measured with W1W2, S2W2 and E2S2 CHARA-telescope-pairs  are close to and just prior to the first minimum in the visibility curve (see Fig \ref{MWC275_Vis_CHARA}). In this region the visibility-baseline relation  for the emission models in Fig \ref{rim_atm} can be approximated with a linear function. We calculated reduced $\chi^{2}$ values for the best-fit line to the W1W2, S2W2 and E2S2 visibilities as a function of effective baseline, varying the assumed inclination and position angle of the observed disk. Fig \ref{chi2} shows the reduced $\chi^2$ surface for the fits, plotted against the assumed disk-inclination and position angle. To further illustrate the change in quality of fits as inclination and position angles are varied , Fig \ref{PA_inc} shows the linear fits to the visibility vs  effective baseline data set.

As seen in Figs \ref{chi2} and \ref{PA_inc}, the quality of the fits show dramatic improvement at MWC275 disk PA of 136$^{o}{\pm}$2$^{o}$  and inclination of 48$^{o}{\pm}$2$^{o}$.
These values are very close to a disk  PA of 139$^{o}{\pm}$15$^{o}$ and inclination of 51$^{+11}_{-9}$ degrees  determined in \cite{Wassel}. The excellent agreement in inclination and PA values for MWC275 from two independent methods strongly supports a disk model for MWC275,  validating the use of ``effective baselines'' to plot MWC275 visibilities.

\clearpage

\bibliographystyle{apj}
\bibliography{model_temp}

\clearpage


%
%



%









\clearpage
\end{document}